\PassOptionsToPackage{x11names, svgnames, dvipsnames, table}{xcolor}
\documentclass[letterpaper,twocolumn,10pt]{article}

\usepackage{usenix-2020-09}
\usepackage{graphicx}
\usepackage{booktabs}

\usepackage{tikz}
\usetikzlibrary{shapes, arrows.meta, positioning, calc, backgrounds, fit, patterns, decorations.pathreplacing}

\usepackage{pgfplots}
\usepgfplotslibrary{colormaps,groupplots}
\pgfplotsset{compat=1.18}

\definecolor{customblue}{HTML}{DAE8FC}
\definecolor{customblueoutline}{HTML}{7998C6}
\definecolor{custompink}{HTML}{E1D5E7}
\definecolor{custompinkoutline}{HTML}{9673A6}
\definecolor{custombatchbackground}{HTML}{e2e2df}
\definecolor{customorange}{HTML}{FFE6CC}
\definecolor{customorangeoutline}{HTML}{D9A110}
\definecolor{plotorange}{HTML}{E89575}
\definecolor{plotblue}{HTML}{95A3C3}
\definecolor{plotgreen}{HTML}{A3C766}
\definecolor{plotpink}{HTML}{DB95C0}
\definecolor{scattergreen}{HTML}{009E73}
\definecolor{outlinegrey}{HTML}{BFBFBF}

\definecolor{blue1}{HTML}{B3D9FF}
\definecolor{blue2}{HTML}{85B7FF}
\definecolor{blue3}{HTML}{2B82FF}
\definecolor{blue4}{HTML}{005FCC}
\definecolor{green1}{HTML}{CDFF9A}
\definecolor{green2}{HTML}{B4FF66}
\definecolor{green3}{HTML}{65CC01}
\definecolor{green4}{HTML}{4E9900}
\definecolor{orange1}{HTML}{FFD9A9}
\definecolor{orange2}{HTML}{FFBF7F}
\definecolor{orange3}{HTML}{FF8F3F}
\definecolor{orange4}{HTML}{DD6A2B}
\definecolor{barblue}{HTML}{166D9C}
\definecolor{baryellow}{HTML}{C48820}

\definecolor{pastelw1}{HTML}{ff99c8}
\definecolor{pastelw2}{HTML}{fec8c3}
\definecolor{pastelw3}{HTML}{fcf6bd}
\definecolor{pastelw4}{HTML}{d0f4de}
\definecolor{pastelw1b}{HTML}{f375b9}
\definecolor{pastelw1c}{HTML}{e956ab}
\definecolor{pastelw1d}{HTML}{de369d}

\definecolor{pastelgreen}{HTML}{C1E1C1}
\definecolor{eerampcolor}{HTML}{e4c1f9}
\definecolor{layercolor}{HTML}{a9def9}
\definecolor{heldcolor}{HTML}{f6e3a1}

\colorlet{sysflexit}{barblue}
\colorlet{syshold}{orange4}
\definecolor{sysperrow}{HTML}{009E73}
\definecolor{syswithin}{HTML}{8493BA}
\definecolor{syseagle}{HTML}{B4589A}
\colorlet{sysnonee}{black!70}
\colorlet{sysgrouped}{black!55}
\pgfplotsset{
  flexit axis/.style={ymajorgrids=true, grid style={dashed, gray!30}, title style={font=\scriptsize, yshift=-4pt},
    every axis plot/.append style={line width=0.7pt}},
  flexit legend/.style={legend cell align=left,
    legend style={font=\scriptsize, draw=none, /tikz/every even column/.append style={column sep=1.2ex}}},
  sys flexit/.style={color=sysflexit, mark=*, mark size=1.5pt},
  sys hold/.style={color=syshold, mark=square*, mark size=1.4pt},
  sys perrow/.style={color=sysperrow, mark=x, mark size=1.9pt},
  sys within/.style={color=syswithin, mark=triangle*, mark size=1.7pt},
  sys eagle/.style={color=syseagle, mark=diamond*, mark size=1.9pt},
  sys nonee/.style={color=sysnonee, mark=pentagon*, mark size=1.6pt},
  bar flexit/.style={fill=sysflexit, draw=none},
  bar hold/.style={fill=syshold, draw=none},
  bar perrow/.style={fill=sysperrow, draw=none},
  bar within/.style={fill=syswithin, draw=none},
}
\tikzset{nonee ref/.style={black!55, densely dashed, line width=0.5pt}}

\usepackage{xspace}
\usepackage{balance}
\usepackage{adjustbox}
\usepackage{comment}
\usepackage{subcaption}
\usepackage{float}
\usepackage{amsmath}
\usepackage{amssymb}
\usepackage{amsthm}
\usepackage[capitalise,noabbrev, nameinlink]{cleveref}
\usepackage{calc}
\usepackage{wasysym}

\crefname{algorithm}{Algorithm}{Algorithms}
\Crefname{algorithm}{Algorithm}{Algorithms}
\crefformat{section}{#2\S#1#3}
\Crefname{section}{Section}{Section}
\crefname{figure}{Figure}{Figures}
\Crefname{figure}{Figure}{Figures}
\crefname{equation}{Equation}{Equations}
\Crefname{equation}{Equation}{Equations}
\crefname{listing}{Listing}{Listings}
\Crefname{listing}{Listing}{Listings}
\crefname{defn}{definition}{definitions}
\theoremstyle{plain}
\newtheorem{theorem}{Theorem}

\newtheorem{proposition}{Proposition}

\theoremstyle{remark}
\newtheorem*{remark}{Remark}
\crefname{theorem}{Theorem}{Theorems}
\Crefname{theorem}{Theorem}{Theorems}
\crefname{lemma}{Lemma}{Lemmas}
\Crefname{lemma}{Lemma}{Lemmas}
\crefname{proposition}{Proposition}{Propositions}
\Crefname{proposition}{Proposition}{Propositions}
\crefname{corollary}{Corollary}{Corollaries}
\Crefname{corollary}{Corollary}{Corollaries}
\crefname{table}{Table}{Tables}
\Crefname{table}{Table}{Tables}
\crefname{appendix}{Appendix}{Appendices}
\Crefname{appendix}{Appendix}{Appendices}

\usepackage{tabularx}
\usepackage[inline, shortlabels]{enumitem}

\usepackage{multirow}
\usepackage[colorinlistoftodos]{todonotes}

\newcommand{\sys}{\textsc{DREX}\xspace}
\newcommand{\syshold}{\sys(holding only)\xspace}
\newcommand{\syscphold}{\sys(compacted, holding only)\xspace}

\definecolor{pennred}{RGB}{153,0,0}

\newcommand{\topheading}[1]{\noindent\textbf{#1.}}
\newcommand{\heading}[1]{\vspace{4pt}\noindent\textbf{#1.}}

\newenvironment{vinlist}
{\begin{itemize}[leftmargin=1.5em, topsep=0pt, partopsep=0pt, itemsep=0pt, parsep=0pt]
  \setlength{\itemsep}{0pt}
  \setlength{\labelwidth}{0.75em}
  \setlength{\parsep}{0pt}
  \setlength{\parskip}{2pt}
  \setlength{\topsep}{0pt}
  \setlength{\partopsep}{0pt}
  }
{\end{itemize}}

\newenvironment{vinenum}
{\begin{enumerate}[leftmargin=1.75em, topsep=0pt, partopsep=0pt, itemsep=0pt, parsep=0pt]
  \setlength{\itemsep}{0pt}
  \setlength{\labelwidth}{1em}
  \setlength{\parsep}{0pt}
  \setlength{\parskip}{2pt}
  \setlength{\topsep}{0pt}
  \setlength{\partopsep}{0pt}
  }
{\end{enumerate}}

\makeatletter
\newcommand{\crefnames}[3]{%
  \@for\next:=#1\do{%
    \expandafter\crefname\expandafter{\next}{#2}{#3}%
  }%
}
\makeatother

\DeclareRobustCommand*\circled[1]{\tikz[baseline=(char.base)]{
    \node[shape=circle,draw,inner sep=0.75pt] (char) {\footnotesize #1};}}

\crefnames{part,chapter,section}{\S}{\S\S}

\newcommand{\peff}{p_{\mathrm{eff}}}
\newcommand{\Gone}{G_1}
\newcommand{\Bd}{B_d}
\newcommand{\Db}{D_b}

\makeatletter
\renewcommand{\@maketitle}{\newpage
 \vbox to 1.5in{
 \vskip 1em
 \begin{center}%
  {\Large\bf \@title \par}%
  \vskip 0.375in minus 0.300in
  {\large\it
   \lineskip 0em
   \renewcommand{\arraystretch}{0.9}
   \begin{tabular}[t]{c}\@author
   \end{tabular}\par}%
 \end{center}%
 \par
 }
}
\makeatother

\begin{document}

\date{}

\title{Batched Early-Exit LLM Inference Done Right with \sys}

\author{
{\rm Xuting Liu\textsuperscript{1}, Daniel Alexander\textsuperscript{2}, Jiaheng Lu\textsuperscript{1}, Siva Kesava Reddy Kakarla\textsuperscript{3},}\\
{\rm Behnaz Arzani\textsuperscript{3}, Vincent Liu\textsuperscript{1}}\\
\textsuperscript{1}University of Pennsylvania, \textsuperscript{2}Purdue University, \textsuperscript{3}Microsoft Research
}

\maketitle

\begin{abstract}
Early-exit (EE) LLMs let confident tokens skip a model's deep layers, but batched serving loses most of this gain. Current systems either force tokens to exit before their own ramp is confident, an error the ramp does not bound, or save almost nothing when decoding is weight-bound, because any continuing token makes a step read the deep weights.
We present \sys, a batched EE serving system that never forces a token to exit. 
\sys introduces cross-iteration batching: tokens that continue past the exit layer wait there, and every few steps one deep pass over all waiting tokens reads the deep weights once for several steps.
Its two policies differ in what such a token's request does: holding pauses it until the deep pass, and speculative holding lets it keep decoding from the exit layer's own guess, which the next deep pass verifies.
Exited tokens are trusted and never verified.
Using a calibrated cost model, a controller picks speculative holding where reading weights dominates a step and guesses are often right, and holding elsewhere.
\sys keeps vLLM's asynchronous scheduling by planning each deep pass from exit counts of earlier steps.
With synthetic exits where 90\% of tokens exit at a quarter of the depth and 40\% of guesses are correct, \sys achieves 2.22--2.24$\times$ decode speedup over non-EE across batch sizes on Llama-2-70B, while per-row skipping (vSkipper's design) gains at most 7.0\%.
On LayerSkip models, \sys adds no statistically resolved quality loss to the EE model's own.
On CNN/DM at full memory, \sys with LayerSkip achieves 1.58--1.91$\times$ decode speedup while remaining within 1.4 ROUGE-2 points of the non-EE model.
These speedups are 9--63\% larger than EAGLE's across the evaluated model sizes, with each method measured against its own non-EE target.

\end{abstract}

\section{Introduction}
\label{sec:intro}
Early-exit (EE) LLMs let a token whose prediction is already confident skip a model's deep layers~\cite{calm2022,bae2023free,elhoushi2024layerskip,fan2024adainfer,tang2024deed} (\cref{fig:earlyexit}).
For a single request, if 90\% of its decoded tokens exit after a quarter of the layers, end-to-end time is reduced by 3.1$\times$.
Unfortunately, requests are rarely processed in isolation~\cite{orca,pagedattention}, and existing EE systems handle batching poorly.

The primary issue is that existing policies force a tradeoff: either the whole batch exits or they all continue together~\cite{miao2024efficient}.
This frequently forces tokens to make decisions contrary to their own ramp's advice, in what we call \emph{involuntary exits} and \emph{involuntary stays}.
Involuntary exits corrupt the context of every later token: on a LayerSkip Llama-2-13B model~\cite{elhoushi2024layerskip}, majority voting forces 17--28\% of tokens out and loses 30 points of BBH accuracy.
Involuntary stays do not hurt quality, because the token runs the full model, but handled naively it gives up the throughput gain---batches rarely exit and performance is close to non-EE.

The obvious answer is to split the batch so that each token exits or continues on its own, but tokens that continue are confronted with yet another dilemma.
If we let the tokens proceed to the deeper layers, the step still reads all of the deep weights, and at moderate batch sizes where a decode step is bound by weight reads, the exits save almost nothing.
If, instead, we hold them until enough other requests have also continued, one deep pass can serve many waiting tokens, but the waiting requests are blocked, which costs memory and TPOT.
With limited KV cache capacity, the shallow layers will run over fewer rows while requests wait, reducing the useful amount of work per weight read.

We argue that batched EE serving should adhere to:
\begin{vinenum}
    \item \textit{No forced exits:} batching never makes a token exit against its own EE ramp decision.
    \item \textit{Fill both passes:} subject to the first principle, each read of the shallow/deep weights serves as many tokens as the batch and the KV-cache budget allow.
\end{vinenum}

\suppressfloats[t]
\begin{figure}[t]
\centering
\begin{minipage}[c]{0.5\linewidth}
\centering
\scalebox{0.72}{\usetikzlibrary{backgrounds, positioning, shapes.geometric, arrows.meta}
\definecolor{eeshallow}{HTML}{D4F1F9}
\definecolor{eedeepbg}{HTML}{B3BDF3}
\definecolor{eedeep}{HTML}{7F8FE8}

\begin{tikzpicture}[
  node distance=0.3cm, font=\small,
  scale=1.0,
  every node/.style={draw, minimum width=0.5cm, minimum height=0.5cm, align=center},
  arrow/.style={thick,-Stealth}
]

  \pgfdeclarelayer{background}
  \pgfsetlayers{background,main}

  \node[rounded corners, fill=white] (word1) {The};
  \node[rounded corners, fill=heldcolor, right=of word1] (word2) {sky};
  \node[rounded corners, fill=pastelgreen, right=of word2] (word3) {is};
  \node[rounded corners, fill=heldcolor, right=of word3] (word4) {blue};

  \node[fill=eeshallow, line width=0.8pt,
        minimum width=4cm, below=0.9cm of word2.north, anchor=north, xshift=0.5cm] (bottom) {Shallow Layers};
  
  \node[fill=eerampcolor, line width=0.8pt, diamond, minimum size=0.4cm, aspect=5, below=0.1cm of bottom.south, anchor=north, align=center] (ee) {EE ramp};

  \node[rounded corners, fill=pastelgreen, below=1.2cm of bottom.south west, anchor=north west, xshift=1.25cm] (box2) {is};
  \node[rounded corners, fill=pastelgreen, below=1.2cm of bottom.south west, anchor=north west, xshift=2.95cm] (box4) {EOS};

  \node[fill=eedeepbg, line width=0.8pt,
        minimum width=4cm, below=0.2cm of box2.south west, anchor=north west, xshift=-1.3cm] (bottom2) {Deep Layers};

  \node[rounded corners, fill=heldcolor, below=0.2cm of bottom2.south west, anchor=north west, xshift=0.15cm] (box1) {sky};
  \node[rounded corners, fill=heldcolor, below=0.2cm of bottom2.south west, anchor=north west, xshift=2.1cm] (box3) {blue};

  \begin{pgfonlayer}{background}
    \foreach \word in {word1, word2, word3, word4} {
      \draw[arrow] (\word.south) -- ++(0,-0.4);
    }

    \draw[arrow] 
      ($ (bottom.south -| bottom2.north) + (-1.5cm,0) $) -- 
      ($ (bottom2.north) + (-1.5cm,0) $);
    \draw[arrow] 
      ($ (bottom.south -| bottom2.north) + (0.475cm,0) $) -- 
      ($ (bottom2.north) + (0.475cm,0) $);
    \draw[arrow] (bottom.south -| box2.north) -- (box2.north);
    \draw[arrow] (bottom.south -| box4.north) -- (box4.north);

    \draw[arrow] (box1.east) -| ++(0.215cm,0) -| ++(0,4.7cm) -| ++(0.5cm,0cm) -- ++(0,-0.35cm);
    \draw[arrow] (box3.east) -| ++(0.05cm,0) -| ++(0,4.7cm) -| ++(0.45cm,0cm) -- ++(0,-0.35cm);
    \draw[arrow, line width=1pt, draw=eerampcolor] (box2.east) -| ++(0.25cm,0) -| ++(0,3.245cm) -| ++(0.4cm,0cm) -- ++(0,-0.35cm);

    \draw[thick] (bottom2.south -| box1.north) -- (box1.north);
    \draw[thick] (bottom2.south -| box3.north) -- (box3.north);
  \end{pgfonlayer}

\end{tikzpicture}}
\end{minipage}\hfill
\begin{minipage}[c]{0.47\linewidth}
\caption{Conceptual diagram of an LLM with EE.
Each token is a separate step.
The sequence exits early in the 2nd and 4th steps (when outputting the tokens ``is'' and ``EOS.'')
}
\label{fig:earlyexit}
\end{minipage}
\end{figure}

\sys meets the first principle by construction.
Every token exits exactly when it would in batch-1 EE (the EE model decoding its request alone), except when a more accurate `stay' decision would benefit performance.
The second principle is harder because the two passes compete: to fill the deep pass, continuing tokens must wait for one another, and each waiting request holds KV memory without decoding, which leaves the shallow pass with fewer rows.
We call the rule that decides, at each step, which tokens run which layers a \textit{batched EE policy}.
This paper asks which policy best resolves this tension.

The first insight is that letting continuing tokens wait at the EE ramp reduces waste in the deep pass, because a weight-bound deep pass costs about the same for one token as for many.
We therefore introduce \emph{cross-iteration batching}, which serves the continuing tokens of several steps with one read of the deep weights.
Its first policy, \emph{holding}, pauses each request whose token continues at the exit layer, keeps decoding the others, and every $k$ steps runs one deep pass, a flush, over all held tokens (\cref{fig:schedules}b).
In a weight-bound cost model, holding is near-optimal among policies that pause requests: none exceeds the speedup of a request decoded alone, and a fixed flush period comes within a few percent of the best.

The second insight is that a held request need not pause, because pausing wastes the pass of the shallow layers.
\emph{Speculative holding}, the second policy of cross-iteration batching, combines holding with self-speculative decoding~\cite{elhoushi2024layerskip,zhang2024draftverify}.
When a request's token continues at the EE ramp and would wait for a deep layers pass, \sys takes the exit layer's token as a guess, feeds it to the next shallow layers pass, and lets the next flush verify it (\cref{fig:schedules}c).
With independent exits and guesses, a position then commits with probability $p + (1-p)\alpha$, where $\alpha$ is the guess's accuracy, and a fixed flush period comes within a few percent of the best such policy.
Exited tokens, on the other hand, stay trusted and need no verification.
Once large batches make decoding compute-bound, verifying guessed tokens is no longer free, so pure speculative decoding's speedup falls while the speedup from EE grows. 

\begin{figure}[t]
\centering
\definecolor{eeshallow}{HTML}{D4F1F9}
\definecolor{eedeepbg}{HTML}{B3BDF3}
\definecolor{eedeep}{HTML}{7F8FE8}
\begin{tikzpicture}[font=\scriptsize, x=0.64cm, y=0.34cm]
  \tikzset{
    cell/.style={draw=black!60, line width=0.3pt, minimum width=0.36cm, minimum height=0.27cm, inner sep=0pt},
    ex/.style={cell, fill=pastelgreen},
    co/.style={cell, fill=heldcolor},
    tn/.style={densely dashed, draw=black},
    wt/.style={cell, fill=black!7, draw=black!25},
    dr/.style={cell, fill=eedeep},
    sblk/.style={fill=eeshallow, draw=black!45, line width=0.3pt},
    dblk/.style={fill=eedeepbg, draw=black!45, line width=0.3pt},
    blab/.style={rotate=90, text=black!65, inner sep=0pt},
    rl/.style={anchor=east, font=\scriptsize},
    pt/.style={anchor=west, inner sep=0pt, font=\scriptsize\bfseries},
    disc/.style={red!80!black, line width=0.6pt},
  }
  \newcommand{\strike}[2]{\draw[disc] ($(#1,#2)+(-0.17cm,-0.11cm)$) -- ($(#1,#2)+(0.17cm,0.11cm)$);}
  \newcommand{\shallowbox}[1]{\draw[sblk] (#1+0.03,-0.3) rectangle (#1+0.97,-4.7); \node[blab] at (#1+0.18,-2.5) {Shallow};}
  \newcommand{\deepbox}[1]{\draw[dblk] (#1+0.03,-0.5) rectangle (#1+2.17,-4.5); \node[blab] at (#1+1.97,-2.5) {Deep};}
  \newcommand{\rowlabels}[1]{\foreach \r in {1,...,4} {\node[rl] at (#1-0.05,-\r) {r\r};}}

  \node[pt] at (-0.75,0.55) {(a) Within-iteration batching};
  \draw[-Stealth, black!60] (6.6,0.55) -- (9.6,0.55) node[anchor=south east, inner sep=1pt, yshift=1.5pt] {time};
  \foreach \s in {0,1,2} {\shallowbox{3.2*\s} \deepbox{3.2*\s+1}}
  \rowlabels{0}
  \foreach \s/\a/\b/\c/\d in {0/ex/co/ex/ex, 1/ex/co/ex/co, 2/ex/ex/co/ex} {
    \pgfmathtruncatemacro{\p}{\s+1}
    \node[\a] at (3.2*\s+0.66,-1) {\p}; \node[\b] at (3.2*\s+0.66,-2) {\p};
    \node[\c] at (3.2*\s+0.66,-3) {\p}; \node[\d] at (3.2*\s+0.66,-4) {\p};
  }
  \node[dr] at (1.75,-2) {1};
  \node[dr] at (4.95,-2) {2}; \node[dr] at (4.95,-4) {2};
  \node[dr] at (8.15,-3) {3};

  \begin{scope}[yshift=-6.3*0.34cm]
    \node[pt] at (-0.75,0.55) {(b) Holding};
    \foreach \i in {0,1,2} {\shallowbox{\i}}
    \deepbox{3}
    \rowlabels{0}
    \node[ex] at (0.66,-1) {1}; \node[ex] at (1.66,-1) {2}; \node[ex] at (2.66,-1) {3};
    \node[co] at (0.66,-2) {1}; \node[wt] at (1.66,-2) {};  \node[wt] at (2.66,-2) {};
    \node[ex] at (0.66,-3) {1}; \node[ex] at (1.66,-3) {2}; \node[co] at (2.66,-3) {3};
    \node[ex] at (0.66,-4) {1}; \node[co] at (1.66,-4) {2}; \node[wt] at (2.66,-4) {};
    \node[dr] at (3.75,-2) {1}; \node[dr] at (3.75,-3) {3}; \node[dr] at (3.75,-4) {2};
  \end{scope}

  \begin{scope}[yshift=-6.3*0.34cm, xshift=6.6*0.64cm]
    \node[pt] at (-0.75,0.55) {(c) Speculative holding};
    \foreach \i in {0,1,2} {\shallowbox{\i}}
    \deepbox{3}
    \rowlabels{0}
    \node[ex] at (0.66,-1) {1}; \node[ex] at (1.66,-1) {2}; \node[ex] at (2.66,-1) {3};
    \node[co] at (0.66,-2) {1}; \node[co, tn] at (1.66,-2) {2}; \node[ex, tn] at (2.66,-2) {3};
    \node[ex] at (0.66,-3) {1}; \node[ex] at (1.66,-3) {2}; \node[co] at (2.66,-3) {3};
    \node[ex] at (0.66,-4) {1}; \node[co] at (1.66,-4) {2}; \node[ex, tn] at (2.66,-4) {3};
    \node[dr] at (3.75,-2) {1}; \node[dr] at (4.4,-2) {2};
    \node[dr] at (3.75,-3) {3}; \node[dr] at (3.75,-4) {2};
    \node[text=red!80!black, font=\footnotesize] at (3.27,-2) {$\times$};
    \node[text=green4, font=\footnotesize] at (3.27,-4) {\checkmark};
    \strike{1.66}{-2} \strike{2.66}{-2} \strike{4.4}{-2}
  \end{scope}

  \begin{scope}[yshift=-12.1*0.34cm]
    \node[ex] at (-0.4,0) {}; \node[anchor=west] at (-0.1,0) {exit};
    \node[co] at (1.6,0) {}; \node[anchor=west] at (1.9,0) {continue};
    \node[cell, tn, fill=white] at (4.3,0) {}; \node[anchor=west] at (4.6,0) {tentative};
    \node[wt] at (7.0,0) {}; \node[anchor=west] at (7.3,0) {paused};
    \node[dr] at (-0.4,-1.15) {}; \node[anchor=west] at (-0.1,-1.15) {deep row};
    \draw[disc] (2.1,-1.32) -- (2.46,-0.98); \node[anchor=west] at (2.6,-1.15) {discarded};
    \node[text=green4, font=\footnotesize] at (4.95,-1.15) {\checkmark}; \node[anchor=west] at (5.15,-1.15) {accepted guess};
    \node[text=red!80!black, font=\footnotesize] at (8.45,-1.15) {$\times$}; \node[anchor=west] at (8.65,-1.15) {rejected guess};
  \end{scope}
\end{tikzpicture}
\caption{Three faithful policies for the same requests and exit decisions (numbers are positions). Blue cells are the tokens a deep read serves: (a)~wastes most of each read, (b)~fills it by pausing requests, and (c)~fills it further by letting held requests keep decoding from the exit layer's token, which the read verifies.
}
\label{fig:schedules}
\end{figure}

No one policy is best everywhere: speculative holding pays where reading weights dominates a step and guesses have a high acceptance rate, and holding wins elsewhere.
\sys implements both policies, and a controller picks the policy from a calibrated cost model.
\sys also plans whether each step runs a deep pass from exited token counts, thereby keeps vLLM's asynchronous scheduling. 

\sys is a serving system, not an EE model, and it sees a model only through its exit depth, exit rate, and guess acceptance.
We therefore evaluate it mainly with a synthetic ramp that sweeps all three on real models.
When 90\% of tokens exit at a quarter of the depth and 40\% of guesses are right, \sys reaches 2.22--2.24$\times$ throughput of non-EE serving on Llama-2-70B(\cref{fig:map}a), and an EE model with this exit pattern on Llama-3.1-8B would beat EAGLE-3 once the batch is large (\cref{fig:crossover}).
Today's EE models, such as LayerSkip, exit this often only at a large loss in quality.
That is a limitation of the models, not of their serving: \sys adds no loss of its own, so as EE models improve, \sys turns their exits into higher throughput at their own quality.

To the best of our knowledge, \sys is the first batched EE serving system that turns exits into throughput without forcing any token to exit, and the first to beat, by speculating, the best policy without speculation.
Its contributions are:
\begin{vinlist}
    \item Cross-iteration batching with two faithful policies, holding and speculative holding, and a theory of faithful batched EE from the batch-1 bound to the optimal full-flush policy, which a controller uses to pick between them (\cref{sec:theory}).
    \item \sys, with a compacted deep pass, coupled guessing, and plan-ahead flushing built in vLLM (\cref{sec:design}).
    \item A model-agnostic evaluation over exit depth, rate, and acceptance, and a cost model on when \sys beats speculative decoding (EAGLE) (\cref{sec:eval}).
\end{vinlist}

\section{Background and Motivation}
\label{sec:background}

\subsection{Batched LLM Decoding}
\label{sec:background-decode}

After prefill processes a request's prompt, decode produces one token per step~\cite{attention}.
A couple of optimizations are particularly critical to making repeated passes through the model efficient.
The first is KV caching in the attention layers, which stores/reuses the key and value tensors computed for every previously processed token.
The second is continuous batching~\cite{orca}, which makes scheduling decisions at the granularity of individual decode steps rather than whole requests, enabling more opportunities for batching and amortization of model weight loads.

In the end, however, at moderate batch sizes and short contexts, a batched decode step is often bounded by model weight I/O.
For Llama-2-70B on a GB200, for instance, decoding takes 25.4\,ms with 16 requests and 28.1\,ms with 47 (2.9$\times$ the tokens cost only 1.10$\times$ the time).

\subsection{Early-Exit Models}
\label{sec:background-ee}

EE models allow some tokens to exit before the last layer~\cite{calm2022,bae2023free,fan2024adainfer,men2025shortgpt,jamialahmadi2025balcony,skipdecode,eellm,pan2024eetuning,tang2024deed,elhoushi2024layerskip}, while layer-skipping models let a token bypass individual layers~\cite{mod2024,dllm2024}.
In an EE model, a ramp at an intermediate layer decides whether the hidden state at that layer is sufficient to produce the token.
The ramp can be a small classifier~\cite{calm2022,apparate2024} or, as in LayerSkip~\cite{elhoushi2024layerskip}, the model's own LM head applied at the exit layer.

Early exits leave a gap in the KV cache.
A later token that continues into the deep layers must attend to deep-layer KV entries for earlier tokens that exited.
Recent systems like CALM~\cite{calm2022,Elbayad2020Depth-Adaptive,miao2024efficient} fill in these entries from the exit layer's hidden state.
We treat all of these mechanisms---the ramp, its threshold, and state propagation---as the EE model's semantics, and build a serving system around them.

\subsection{Grouped Policies Force Exits}
\label{sec:background-grouped}

In a batch, some tokens of a step want to exit while others want to continue.
Grouped policies resolve this split by making the step exit or continue as a whole~\cite{miao2024efficient}, and we compare three such rules: Consensus exits only if every token agrees, Greedy if any token wants to, and Majority if more than half do.
\Cref{tab:involuntary} measures them on LayerSkip Llama-2-13B against two references: non-EE, which runs every token through all layers, and batch-1 EE, which decodes each request alone and so gives the EE model's own accuracy.
Every grouped rule must force tokens to exit early or to give up their exit: Majority and Greedy, which force exits, lose most of the EE model's accuracy, while Consensus, which forces only stays, scores above batch-1 EE but almost never exits, so at full memory its throughput is no higher than non-EE even without scheduling overhead (\cref{sec:appendix-grouped}).

\begin{table}[t]
\caption{Exits, involuntary exits and stays, and accuracy (exact match) on LayerSkip Llama-2-13B (exit after layer 15 of 40, 50\% exit-rate target, $N = 64$, held-out documents).
}
\label{tab:involuntary}
\centering
\scriptsize
\setlength{\tabcolsep}{2.2pt}
\begin{tabularx}{\columnwidth}{X cc cc cc cc}
\toprule
\multirow{2}{*}{\textbf{Policy}} & \multicolumn{2}{c}{\textbf{Exits (\%)}} & \multicolumn{2}{c}{\textbf{Invol. exit (\%)}} & \multicolumn{2}{c}{\textbf{Invol. stay (\%)}} & \multicolumn{2}{c}{\textbf{Accuracy (\%)}} \\
\cmidrule(lr){2-3}\cmidrule(lr){4-5}\cmidrule(lr){6-7}\cmidrule(lr){8-9}
 & \textbf{GSM8K} & \textbf{BBH} & \textbf{GSM8K} & \textbf{BBH} & \textbf{GSM8K} & \textbf{BBH} & \textbf{GSM8K} & \textbf{BBH} \\ \midrule
Non-EE & -- & -- & -- & -- & -- & -- & 26.4 & 46.8 \\
Batch-1 EE & 50.7 & 50.1 & -- & -- & -- & -- & 21.6 & 43.0 \\
\midrule
Consensus & 0.5 & 0.1 & 0 & 0 & 49.1 & 49.6 & 25.2 & 46.2 \\
Majority & 83.5 & 91.9 & 17.0 & 28.4 & 7.3 & 3.6 & 3.2 & 13.2 \\
Greedy & 100 & 100 & 20.0 & 29.8 & 0 & 0 & 2.0 & 8.8 \\
\midrule
\sys & 53 & 50 & 0 & 0 & 0 & 0 & 20.8 & 44.0 \\
\bottomrule
\end{tabularx}
\end{table}

\subsection{Exits Within a Step Save Little}
\label{sec:background-opportunity}

Honoring every token's own decision removes involuntary choices but does not turn exits into throughput.
In a batch, all $N$ tokens of a step exit with probability $p^N$, so the deep layers read their weights in almost every step.
Within-iteration batching, which skips the deep layers only for a step's exits, and per-row skipping~\cite{vskipper2026}, which compacts each step's deep pass to its continuing tokens, therefore save only per-token work such as attention and row compute.
These savings are real where attention or row compute dominate a step, but where weight reads bound it, each read of the deep weights must serve the continuing tokens of several iterations.

\section{Cross-Iteration Batching}
\label{sec:theory}

Cross-iteration batching runs the deep layers once for the continuing tokens of several steps, so one pass of the deep layers serves them all.
Its two policies differ in what a request whose token continues does until that deep pass: holding pauses it, and speculative holding lets it keep decoding from a guess.

How much throughput can a faithful policy reach, and when does each policy pay?
We answer both in a weight-bound model W, where a step's time is set by the layers it reads, so filling both passes raises throughput (proofs in \cref{sec:appendix-proofs}).
A calibrated W guides the controller on \sys.

\subsection{Faithfulness and the Weight-Bound Model}
\label{sec:theory-model}

\topheading{The EE model and faithfulness}
An EE model has $L$ layers and one EE ramp after layer $E$, and $r = E/L$ is its exit depth.
At each position, the shallow layers $1..E$ produce the hidden state $h_E$, and the ramp decides whether the position exits with the exit layer's token or continues through the deep layers $E{+}1..L$ to their token $y$.
The exit rate $p$ is the probability that the ramp says exit, which in the long run is the fraction of decode tokens that exit.

A batched EE policy decides which tokens run which layers at each step.
A policy is \emph{faithful} if every request emits exactly the tokens the EE model emits for it alone, which we call batch-1 EE: the same tokens under greedy decoding, and the same tokens under the same per-position noise when sampled.
Faithfulness carries batch invariance~\cite{thinkingmachines2025} to exit decisions: no token exits or stays involuntarily, and grouped policies are not faithful.
Principle 1 forbids only involuntary exits, but in W below an involuntary stay never shortens a critical path (Lemma 2 in \cref{sec:appendix-proofs}), so \cref{thm:batch1,thm:hold} also bound policies that force stays, and the theory loses nothing by studying faithful ones.

\heading{The weight-bound model}
$N$ requests are resident at equal memory.
Steps run one after another, and each reads every layer at most once, in order, over any set of rows.
Every step stays below the ridge $\theta^*$, where rows are free, so a step that reads a set of layers $\Lambda$ takes $|\Lambda| \cdot t_w$, where $t_w$ is the time to read one layer's weights.
A position can start in the step after its predecessor's token is determined, and exits are i.i.d.\ with probability $p$, as in our synthetic ramp.
Speedup is throughput over decoding without EE, and what EE promises is the batch-1 gain, the speedup of one request decoded alone:
\begin{equation}
\Gone = \frac{L}{pE + (1-p)L} = \frac{1}{1 - p(1-r)},
\label{eq:g1}
\end{equation}
which is 3.1$\times$ at $(E/L, p) = (.25, .9)$.

\subsection{When Requests Pause, Holding Is Near-Optimal}
\label{sec:theory-hold}

A policy is \emph{non-speculative} if it never runs a layer for a position whose context holds an undetermined token, a continuing token whose deep pass has not run.

\begin{proposition}[Batch-1 bound]
\label{thm:batch1}
In W, every non-speculative faithful policy has speedup at most $\Gone$, where $p$ is the long-run exit fraction, on every path in the long run.
\end{proposition}

\heading{Holding}
Holding pauses a request whose token continues: it holds every token whose ramp says continue at the exit layer and, every $k$ steps, runs one pass of the deep layers, a flush, over all held tokens.
We call $k$ the flush period, a step that runs only the shallow layers a shallow step, and one that also runs the deep layers a flush step.
In the $i$-th step of a cycle, only the $p^{i-1}N$ requests that exited in every earlier step are ready, so a cycle of $k-1$ shallow steps and one flush step emits $N(1-p^k)/(1-p)$ tokens in $\big((k-1)E + L\big) t_w$, a speedup of
\begin{equation}
S(k; p, r) = \frac{1-p^{k}}{1-p} \cdot \frac{1}{1 + (k-1)r}.
\label{eq:speedup}
\end{equation}
This is the speedup of speculative decoding~\cite{speculative2023} with acceptance $p$, $k-1$ drafts, and draft cost $r$, where exits are accepted without verification.
It is below $\Gone$ for every $k$, and beats within-iteration batching ($k = 1$) only if $p > E/L$.

\begin{theorem}[Holding optimum]
\label{thm:hold}
With i.i.d.\ exits, non-speculative faithful policies form a Markov decision process over the number of ready requests, with optimal speedup $S^*_N$.
The optimal policy runs a deep-only flush, a flush that skips the shallow layers, once fewer than $\lambda^* E$ requests are ready, where $\lambda^* = (N/L)\,S^*_N$ is the optimal rate in tokens per layer read (proof in \cref{sec:appendix-proofs}).
\end{theorem}

Where holding pays ($p > E/L$), fixed flush periods come within 1.7--2.8\% of $S^*_N$ at $N = 16$ and 0.4--0.8\% at 64 (\cref{tab:optimum}), so the controller picks only $k$.
Exceeding $S^*_N$ therefore requires speculation.

\subsection{Guesses Raise the Effective Exit Rate}
\label{sec:theory-spec}

Speculative holding keeps a held request decoding in a self-speculative style~\cite{elhoushi2024layerskip,zhang2024draftverify} instead of pausing it.
The exit layer's own token at the held position is fed as a guess, later positions are tentative, and the next flush keeps the tentative tokens after each correct guess and rolls back from the first wrong one.
Exits are trusted and never verified, so speculative holding stays faithful.
The acceptance $\alpha$, the probability that the guess equals the deep token at a continuing position, is a property of the EE model, which our synthetic ramp sweeps like $p$.

\begin{proposition}[Acceptance]
\label{thm:acceptance}
In any single-guess policy (one guess per held position) whose decisions depend neither on guess outcomes nor on signals correlated with them, a position decoded after $c$ unresolved continuing positions commits with the probability that all $c$ of their guesses are correct.
If exits and guess outcomes are i.i.d., this is $\alpha^c$, and a request that decodes in every step of a window commits position $s = 0, 1, \ldots$ of the window with probability $\peff^{\,s}$, where $\peff = p + (1-p)\alpha$.
\end{proposition}

The acceptance is thus fixed by the model, so a policy only chooses when to flush.

\begin{remark}
Under greedy decoding, $\peff = (1 - e) + w$, where $1 - e$ is the acceptance of self-speculation from the same layer~\cite{elhoushi2024layerskip} and $w$ the probability that a position exits with a token that differs from the deep one.
This is the greedy, same-layer case of Lemma 3 of speculative cascades~\cite{narasimhan2025cascades}, so trusting exits never accepts less than same-layer self-speculation.
\end{remark}

We call a single-guess policy a \emph{full-flush policy} if each step reads the shallow layers over some of the requests that can advance, the deep layers over every unresolved continuing position, or both, as holding and fixed windows do.

\begin{theorem}[Best full-flush policy]
\label{thm:spec}
With i.i.d.\ exits and guesses, full-flush policies form a Markov decision process over the histogram $(n_c)$, where $n_c$ requests have $c$ unresolved continuing positions.
(a)~A fixed window of $k$ steps achieves exactly $S(k; \peff, r)$ while a window's rows stay below the ridge.
(b)~The optimal full-flush policy, of rate $\lambda^*_\alpha$, flushes once a shallow step's expected yield, $\sum_c n_c \alpha^c$, falls below $\lambda^*_\alpha E$ (proof in \cref{sec:appendix-proofs}).
\end{theorem}

In the settings of \cref{tab:optimum} with $\alpha \geq .2$, a fixed window comes within 2.2\% of the best full-flush policy at $N = 16$ and 0.6\% at 64, and within 2.8\% of a guess-outcome oracle (\cref{tab:spec-optimum}).
The controller, which sees exits two steps late, therefore loses little by picking only $k$, except with very few requests (77\% of the best at $N = 1$).

\subsection{When Rows Are Not Free}
\label{sec:theory-beyond}

\heading{Ridges}
A flush of holding runs $N(1-p^k)$ deep rows, a window of speculative holding about $Nk(1-p)$, and same-layer self-speculation $Nk$, so holding reaches the ridge last.
Past the ridge, a deep pass costs time per row, and the best policy compacts it so each token pays only for its own layers, as per-row skipping~\cite{vskipper2026} does at $k = 1$.

\heading{Attention and the hold cap}
A held row skips the deep attention until its flush, but a tentative row pays its attention and LM-head row for $\alpha^c$ expected tokens.
It therefore pays off only if $\alpha^c \bar\tau > a_{\text{row}}$, where $\bar\tau$ is the time per committed token and $a_{\text{row}}$ the row's own cost, so \sys caps a request's held positions at a hold cap $m$, a pause that never helps in W.

\heading{Three regimes}
A step is weight-bound where rows are nearly free, attention-bound where tentative rows pay attention that held rows skip, and compute-bound where the deep pass runs past the ridge.
The ratio of resident KV to weight bytes separates the first two (0.26 for Llama-2-70B and 5.7 for Llama-2-13B at full memory on one GB200), $N$ against $\theta^*$ the third, and a calibrated version of this model predicts each deployment's regime and policy.

\heading{Draft heads}
A draft head~\cite{eagle2024,eagle3} drafts with one layer but verifies all $Nk$ window rows, so in W it matches speculative holding at a break-even acceptance $\beta^*$ below $\peff$ (.64 against .94 on 70B, (.25, .9), $\alpha = .4$), and past the ridge its verify rows cost time while exits remove deep rows, so $\beta^*$ rises with $N$.

\section{\sys Design}
\label{sec:design}

\cref{fig:design} follows one flush step through its components: flush steps run a compacted deep pass over the held positions only, and the engine also keeps an in-place deep pass that runs every row and masks exits.

\begin{figure}[t]
\centering
\definecolor{eeshallow}{HTML}{D4F1F9}
\definecolor{eedeepbg}{HTML}{B3BDF3}
\definecolor{eedeep}{HTML}{7F8FE8}
\begin{tikzpicture}[font=\scriptsize, x=1cm, y=1cm]
  \tikzset{
    cell/.style={draw=black!60, line width=0.3pt, minimum width=0.5cm, minimum height=0.27cm, inner sep=0pt},
    rcell/.style={draw=black!60, line width=0.3pt, minimum width=0.42cm, minimum height=0.27cm, inner sep=0pt},
    ex/.style={cell, fill=pastelgreen},
    co/.style={cell, fill=heldcolor},
    tn/.style={densely dashed, draw=black},
    dr/.style={cell, fill=eedeep},
    rr/.style={rcell, preaction={fill=white}, pattern=north east lines, pattern color=green4!80},
    rl/.style={anchor=east, font=\scriptsize},
    comp/.style={draw=black!70, fill=pastelw4, rounded corners=2pt, align=center, font=\scriptsize,
      inner sep=1.5pt, minimum height=0.5cm},
    lane/.style={font=\scriptsize\bfseries, rotate=90, text=black!70},
    disc/.style={red!80!black, line width=0.6pt},
    mk/.style={circle, draw, fill=white, inner sep=0pt, minimum size=0.3cm, font=\scriptsize},
  }
  \newcommand{\strike}[3]{\draw[disc] ($(#1,#2)+(-#3,-0.11)$) -- ($(#1,#2)+(#3,0.11)$);}
  \def\ya{-0.1} \def\yb{-0.58} \def\yc{-1.06} \def\yd{-1.54}

  \filldraw[fill=pastelw3, draw=black!50, rounded corners=3pt] (-0.05,1.92) rectangle (6.1,2.72);
  \node[lane] at (0.12,2.32) {Host};
  \node[comp, minimum width=1.35cm] (pc) at (1.15,2.32) {Period\\controller};
  \node[comp, minimum width=1.7cm] (pf) at (3.35,2.32) {Plan-ahead\\flushing};
  \node[comp, minimum width=1.1cm] (gc) at (5.3,2.32) {Gap\\clock};
  \draw[-Stealth] (pc) -- node[below=-1pt] {$k, m$} (pf);
  \draw[-Stealth] (gc) -- (pf);

  \filldraw[fill=black!4, draw=black!50, rounded corners=3pt] (-0.05,1.52) rectangle (6.1,-2.25);
  \node[lane] at (0.12,0.62) {GPU};
  \foreach \x/\n in {0.8/1, 1.45/2, 2.1/3} {\node[cell, fill=eeshallow] (st\n) at (\x,1.3) {\n};}
  \node[anchor=east] at (0.55,1.3) {step};
  \node[cell, fill=eedeepbg, minimum width=1.7cm] (fl) at (3.35,1.3) {flush};
  \draw[-Stealth, densely dashed] (st1.north) -- (0.8,1.66) -- (2.8,1.66) -- (2.8,0 |- pf.south);
  \node[above=-2pt] at (1.8,1.66) {counters};
  \draw[-Stealth] (3.85,0 |- pf.south) -- node[mk, pos=0.5] {3} (3.85,0 |- fl.north);

  \node[comp, minimum width=1.8cm] at (1.45,0.62) {Speculative\\windows};
  \node[comp, minimum width=1.7cm] at (3.35,0.62) {Compacted\\deep pass};
  \node[comp, minimum width=1.45cm] at (5.3,0.62) {Verification};

  \foreach \y/\r in {\ya/r1, \yb/r2, \yc/r3, \yd/r4} {\node[rl] at (0.55,\y) {\r};}
  \node[ex] at (0.8,\ya) {1}; \node[ex] at (1.45,\ya) {2}; \node[ex] at (2.1,\ya) {3};
  \node[co] at (0.8,\yb) {1}; \node[co, tn] at (1.45,\yb) {2}; \node[ex, tn] at (2.1,\yb) {3};
  \node[ex] at (0.8,\yc) {1}; \node[ex] at (1.45,\yc) {2}; \node[co] at (2.1,\yc) {3};
  \node[ex] at (0.8,\yd) {1}; \node[co] (g) at (1.45,\yd) {2}; \node[ex, tn] (t) at (2.1,\yd) {3};
  \node[mk] at ($(g.south west)+(-0.02,-0.08)$) {1};
  \draw[-Stealth] (g.south) to[out=-60, in=-120] node[mk, pos=0.5] {2} (t.south);

  \draw[fill=eedeepbg, draw=black!45, line width=0.3pt] (2.5,0.17) rectangle (4.2,-1.8);
  \node[mk] at (2.5,0.17) {4};
  \node[rr] at (2.81,\ya) {1}; \node[rr] at (3.28,\ya) {2}; \node[rr] at (3.75,\ya) {3};
  \node[dr] at (2.85,\yb) {1}; \node[dr] at (3.4,\yb) {2}; \node[rr] at (3.91,\yb) {3};
  \node[dr] at (2.85,\yc) {3}; \node[rr] at (3.36,\yc) {1}; \node[rr] at (3.83,\yc) {2};
  \node[dr] at (2.85,\yd) {2}; \node[rr] at (3.36,\yd) {1}; \node[rr] at (3.83,\yd) {3};
  \strike{1.45}{\yb}{0.22} \strike{2.1}{\yb}{0.22} \strike{3.4}{\yb}{0.22} \strike{3.91}{\yb}{0.19}

  \node[mk] at (4.65,\ya) {5};
  \node[text=red!80!black, font=\footnotesize] at (4.65,\yb) {$\times$}; \node[co] at (5.05,\yb) {1};
  \node[co] at (5.05,\yc) {3};
  \node[text=green4, font=\footnotesize] at (4.65,\yd) {\checkmark}; \node[co] at (5.05,\yd) {2}; \node[ex] at (5.6,\yd) {3};

  \ifdefined\zoomshortlegend
    \node[rr] at (0.4,-2.55) {}; \node[anchor=west] at (0.67,-2.55) {riding row (K/V only)};
  \else
    \node[ex] at (0.4,-2.55) {}; \node[anchor=west] at (0.68,-2.55) {exit};
    \node[co] at (2.25,-2.55) {}; \node[anchor=west] at (2.53,-2.55) {continue};
    \node[cell, tn, fill=white] at (4.3,-2.55) {}; \node[anchor=west] at (4.58,-2.55) {tentative};
    \node[dr] at (0.4,-2.9) {}; \node[anchor=west] at (0.68,-2.9) {deep row};
    \draw[disc] (2.05,-3.0) -- (2.45,-2.8); \node[anchor=west] at (2.53,-2.9) {discarded};
    \node[text=green4, font=\footnotesize] at (4.3,-2.9) {\checkmark}; \node[anchor=west] at (4.58,-2.9) {accepted guess};
    \node[rr] at (0.4,-3.25) {}; \node[anchor=west] at (0.68,-3.25) {riding row (K/V only)};
    \node[text=red!80!black, font=\footnotesize] at (4.3,-3.25) {$\times$}; \node[anchor=west] at (4.58,-3.25) {rejected guess};
  \fi
\end{tikzpicture}
\caption{\sys's components on the flush step for the window of \cref{fig:schedules}c, in compact mode.
\circled{1}~When the EE ramp says continue, the exit layer's token at the held position is its guess, and \circled{2}~the guess is fed as the next input, so later positions are tentative.
\circled{3}~Plan-ahead flushing plans the flush and its buckets from counters two steps old, with $k$ and $m$ from the period controller and the gap clock's bound.
\circled{4}~The compacted deep pass gathers held positions into deep rows and exits whose deep K and V are pending into K/V-only riding rows.
\circled{5}~Verification accepts a guess if it equals the deep token, and each request commits up to its first rejection and discards the rest.
}
\label{fig:design}
\end{figure}

\subsection{Shallow Steps in Place}
\label{sec:design-shallow}

Every running request is a row of every step, as vLLM's asynchronous scheduler already arranges, and rows never move.
A request that can advance runs the shallow layers and the ramp in its row, and one paused until a flush is a replay row with zero attention length and KV writes to a padding slot.
The host sees committed tokens only: a step that commits $j$ tokens for a request reports $1 - j$ rejected tokens, so vLLM's scheduler treats a held token as a rejected speculative token, applies its stop checks to committed tokens, and needs no new request state.

\subsection{Compacted Deep Pass}
\label{sec:design-compact}

Under speculation, a request has several held positions per flush, and one row cannot carry them.
\sys therefore compacts the deep pass into the rows that need it.
Compacting also helps past the ridge, where the masked rows of an in-place deep pass stop being free.

\heading{Windows and the prefix rule}
Each request keeps a window of the positions it has processed since its last flush, each with its exit-layer hidden state $h_E$, its decision, and the exit layer's token.
On a flush step, a device kernel takes prefix sums, in batch order, of each request's held positions and of its exits whose deep K and V are pending.
It resolves every request whose whole window fits into the planned buckets, $\Bd$ deep rows and $\Db$ riding rows, and the requests after the first that does not fit pause until the next flush, where they go first.
An undersized bucket thus costs latency, never correctness, so estimates can size buckets.

\heading{Deep rows and riding rows}
Each held position becomes its own decode row, next to its request's other rows, with its own position, sequence length, and a gathered copy of its request's block-table row.
Causal order needs no masks, because every deep layer writes all K and V before its attention runs.
The resolved windows' exits ride along as K/V-only riding rows in every deep layer's QKV GEMM, which computes their propagated entries where they are first read, while attention and the MLP skip them.
Each $(\Bd, \Db)$ pair has its own CUDA graph, and below the calibrated ridge the buckets are raised to the largest size that costs no extra time, which makes overflows rare (\cref{sec:appendix-tails}).

\subsection{Speculative Windows}
\label{sec:design-spec}

\heading{Guesses, verification, and rollback}
The hold cap $m$ bounds a request's held positions per window.
After a continue with fewer than $m$ held positions, \sys feeds the exit layer's token at that position as the next input, and every later position of the window is tentative, whether it exits or continues.
A request at $m$ held positions pauses until the flush.
The flush resolves the held positions in order: a guess is accepted if it equals the deep token, and at the first rejection \sys commits the window up to that position and the deep token and discards the rest, whose K and V are rewritten before any read.
Exits are never verified, so a flush runs at most $m$ deep rows per request.
With $m = 1$ this is holding, and with $m = k = 1$ per-row skipping.

\heading{Coupled guessing}
Under sampling, faithfulness asks for the same tokens per seed, not only the same distribution.
vLLM samples by Gumbel-max with noise keyed by the request's seed and the position, so \sys draws a guess and the deep token at its position with the same noise and accepts iff the two draws agree.
The committed token is always the EE model's own draw, so outputs equal batch-1 EE's seed for seed, whereas rejection sampling~\cite{speculative2023,chen2023specsampling} would preserve only the distribution (\cref{sec:appendix-proofs} bounds the acceptance).

\subsection{Plan-Ahead Flushing}
\label{sec:design-plan}

An exit decision exists only partway through a step, on the GPU, so a policy that skips the deep layers in the step that decides must branch mid-step, on the host, which synchronizes every step, or with a conditional graph node~\cite{vskipper2026}.
Because a held token can wait, whether a step flushes need not depend on that step's exits.
\sys's host therefore plans each step before it runs: whether it flushes, and its buckets, from per-request rates in device counters that arrive two steps late, times the steps since the last flush, plus a 10\% margin.
The device decides which tokens exit, wait, or join the flush, so \sys keeps vLLM's two steps in flight with full CUDA graphs and needs no host sync, conditional node, or exit predictor~\cite{cdb2026}.

\subsection{Period Controller and Gap Bound}
\label{sec:design-ctl}

\heading{Period controller}
The controller picks the flush period $k$ and the hold cap $m$, and with them the policy ($m = 1$ is holding).
It picks the pair that maximizes throughput in a calibrated cost model at the current exit rate $\hat p$ over the last 512 decisions, acceptance rate $\hat\alpha$, $N$, and context length, and runs the smallest $k$ within 1\% of the best.
It is never told $\alpha$, and after 512 steps without a guess it runs one window at $m = 4$ to keep $\hat\alpha$ current.
The cost model prices a step as the larger of its GPU and host time, adding per layer the larger of the weight read and the rows' compute, plus attention, KV writes, and the LM head (\cref{sec:appendix-costmodel}).
The cost model predicts the synthetic speedups there with a median measured-over-predicted ratio of 0.981--1.000.
Whether a deployment runs the compacted deep pass (compact mode) or the in-place one is chosen once, from the same model at the operating point the operator expects ($N$, $p$, and $\alpha$).
This is determined during the engine initialization.

\heading{Gap-bounded flushes}
A held token waits up to $k-1$ steps, and its request's tentative tokens wait with it.
An operator can bound the gap between a request's tokens.
The host keeps a gap clock, the time since each held request's last token, and flushes once one more step could let the oldest gap exceed the bound, one shallow step early to absorb jitter (\cref{sec:appendix-tails}).
It also flushes when a prefill has waited longer than an optional budget or every row is held, and prefill chunks ride only on flush steps, as in chunked prefill~\cite{sarathi}.

\subsection{Guarantees}
\label{sec:design-guarantees}

\sys guarantees by construction that no token exits unless its own ramp says exit: latency bounds, memory pressure, and bucket overflow trigger flushes or pauses, never exits, and preemption resumes a request at full depth.
It forces a stay only where that speeds up decoding: where a cost model predicts that EE would be slower, \sys lets go the EE opportunity and runs the full-depth model, its full-depth fallback, which costs no accuracy.
Otherwise every token follows its own ramp, so each request's output is batch-1 EE's, up to floating-point near-ties as for any batched engine, for every request that is not preempted.

\section{Implementation}
\label{sec:implementation}

\sys is built on top of vLLM~\cite{pagedattention} 0.30 with its ModelRunnerV2 and asynchronous scheduling support, in about 7,000 lines of Python with 22 Triton kernels.
\sys also builds its cost model in 2,600 lines of Python.
It uses vLLM's extension points (a scheduler subclass, a model-runner subclass, and an EE model) and two core patches of 22 added lines, which add a step variant to the CUDA graph descriptor and return a per-row rejection count and an extra device-to-host channel with each step's outputs.
We will open-source \sys.

\heading{CUDA graphs}
Each padded batch size has one step graph, and the compacted deep pass has its own graphs keyed by $\Bd$ and $\Db$.

\heading{Ramps}
The softmax confidence ramp~\cite{calm2022,eellm} serves models with a shared LM head such as LayerSkip: it applies the final norm and LM head at the exit layer and exits if the gap between the top two softmax probabilities, CALM's softmax response, exceeds the request's threshold $\tau$ ($\tau = \infty$ disables all exits, which reduces \sys to LayerSkip's lossless self-speculation), and its token at a continuing position is the guess.
The synthetic ramp exits a position if a 32-bit MurmurHash3~\cite{murmurhash} of the request seed and position, over $2^{32}$, is below $p$, and accepts a guess if a hash in a second domain is below $\alpha$, so exit depth, rate, and acceptance are free parameters on any model.
Coupled guessing needs vLLM's standard sampler, whose noise is keyed by seed and position, which every system in our runs uses.

\heading{Preemption}
vLLM preempts a request by freeing its KV blocks and resumes it with a prefill over its prompt and output so far.
\sys delivers the request's committed tokens, purges its window so that no tentative token is ever shown, and resumes it at full depth, as all prefill runs, so preemption adds depth but never forces an exit.
Its later tokens may differ from batch-1 EE's, so the equality with batch-1 EE covers requests that are not preempted, and every quality, fidelity, and batch-invariance result here comes from closed-loop runs with zero preemptions.

\section{Evaluation}
\label{sec:eval}

We evaluate \sys on three models whose steps are bound by weights, attention, and compute, with a synthetic ramp that sets any EE model's exit pattern, on LayerSkip models, and against EAGLE draft heads.

\subsection{Setup}
\label{sec:eval-setup}

\heading{Hardware, models, and workloads}
Every run uses BF16 with CUDA graphs and asynchronous scheduling on one GB200 GPU and its Grace NUMA node, and 70B also runs with tensor parallelism 2 on two GPUs of one node.
We serve Llama-2-70B (weight-bound), Llama-2-13B-chat~\cite{llama2} (multi-head attention, attention-bound), Llama-3.1-8B-Instruct~\cite{llama3} (compute-bound at full memory), and the LayerSkip~\cite{elhoushi2024layerskip} Llama-2-13B, Llama-2-70B, and Llama-3-8B checkpoints.
The synthetic ramp sweeps $(E/L, p) \in$ \{(.25, .9), (.375, .7), (.25, .5), (.5, .5), (.75, .3), (.5, 0)\} and $\alpha \in \{0, .2, .4, .6\}$.
Its hash reads only the request seed and the position, every system makes the same decisions for every request.
The cost of a synthetic ramp is set to an LM head pass (0.4\%, 1.3\%, and 6.5\% of 70B's, 13B's, and 8B's weights), matching LayerSkip's mechanism.
Synthetic runs decode a static batch with prompt 200 and output 2000 tokens, at $N = 16$, a middle $N$ (64, 32, and 384 on 13B, 70B, and 8B), and full memory (80, 47, and 527 requests, which shifts with each workload's lengths), with every system pinned to the same KV blocks.
Summarization runs decode held-out CNN/DM~\cite{cnndm} documents at natural length, $N$ at a time, each batch prefilled before it decodes.

\heading{Systems}
Non-EE is default vLLM serving the same model without early exit.
Within-iteration batching honors each exit within its own step, and per-row skipping compacts each step's deep pass to its continuing tokens, which is vSkipper's design~\cite{vskipper2026}, reimplemented in \sys's engine in its best-throughput configuration.
\sys is the whole system, whose controller picks between holding and speculative holding, and each deployment runs the compacted or the in-place deep pass as the cost model picks.
\syshold disables speculation ($m = 1$) and runs in place.
Compact deployments have no fallback, so every LayerSkip result of \sys makes exactly batch-1 EE's decisions.
\syshold runs in place with the full-depth fallback armed, and $\ddagger$ marks results where it ran on at least 1\% of steps (\cref{sec:appendix-layerskip}).
EAGLE is vLLM's lossless speculative decoding with published heads, EAGLE~\cite{eagle2024} for Llama-2-13B-chat and -70B-chat and two EAGLE-3~\cite{eagle3} heads for Llama-3.1-8B, at the best of $k = 1$ to 7 drafted tokens per cell (1 to 4 on summarization).

\heading{Metrics and validity}
We report decode throughput as a speedup over non-EE at the same $N$, the time per output token (TPOT), and the p99 over requests of each request's largest inter-token gap.
Throughput is measured over each run's batch window, from the last request's first token to the first completion (to the last token for summaries), and it includes the controller's warm-up.
Each job runs all its systems on one node, and every closed-loop run is valid: device counters satisfy the commit identities, synthetic decisions equal the host's replay of the hash, and no request is preempted, exits involuntarily, or stays involuntarily outside the fallback.

\subsection{\sys Turns Exits into Throughput}
\label{sec:eval-map}

\begin{figure*}[t]
\centering
\pgfplotsset{bar line legend/.style 2 args={legend image code/.code={
  \draw[fill=#1, draw=none] (0cm,-0.1cm) rectangle (0.25cm,0.1cm);
  \draw[/pgfplots/#2, mark repeat=2, mark phase=2] plot coordinates {(0.3cm,0cm) (0.47cm,0cm) (0.64cm,0cm)};}},
  marker legend/.style={legend image code/.code={\draw[##1] plot coordinates {(0.15cm,0cm)};}}}
\begin{tikzpicture}[font=\footnotesize]
\begin{groupplot}[
  group style={group name=map, group size=3 by 2, horizontal sep=0.9cm, vertical sep=1.1cm},
  scale only axis,
  width=0.262\textwidth, height=1.75cm,
  flexit axis,
  clip=false,
  row a/.style={ybar, /pgf/bar width=4.2pt, ymin=0, ymax=2.7, ytick={0,1,2},
    symbolic x coords={A,B,C,D}, xtick=data, xticklabels={{.25\\.9},{.375\\.7},{.25\\.5},{.5\\.5}},
    x tick label style={font=\scriptsize, align=center}, xtick pos=bottom, enlarge x limits=0.16},
  row b/.style={xmin=-4.3, xmax=1.5, xtick={-4,-3,-2,-1,0,1}, ymin=0.8, ymax=2.25, ytick={1,1.5,2},
    xlabel={ROUGE-2 points vs.\ non-EE}, x tick label style={font=\scriptsize}},
]
\nextgroupplot[row a, ylabel={Speedup},
  legend to name=maplegend, legend columns=5, flexit legend,
  legend image code/.code={\draw[#1, draw=none] (0cm,-0.1cm) rectangle (0.3cm,0.1cm);}]
  \addplot[bar within, bar line legend={syswithin}{sys within}] coordinates {(A,1.038) (B,1.031) (C,1.017) (D,1.019)};
  \addplot[bar perrow, bar line legend={sysperrow}{sys perrow}] coordinates {(A,1.070) (B,1.042) (C,1.032) (D,1.015)};
  \addplot[bar hold, bar line legend={syshold}{sys hold}] coordinates {(A,2.050) (B,1.249) (C,1.198) (D,1.015)};
  \addplot[bar flexit, bar line legend={sysflexit}{sys flexit}, error bars/.cd, y dir=both, y explicit, error bar style={black, line width=0.6pt}, error mark options={black, mark size=1.2pt, rotate=90}]
    coordinates {(A,2.240) += (0,0.209) -= (0,0.136) (B,1.382) += (0,0.128) -= (0,0.081) (C,1.393) += (0,0.228) -= (0,0.151) (D,1.115) += (0,0.076) -= (0,0.065)};
  \draw[nonee ref] ([xshift=-12pt]axis cs:A,1) -- ([xshift=12pt]axis cs:D,1);
  \addlegendentry{within-iteration batching}
  \addlegendentry{per-row skipping}
  \addlegendentry{\sys (holding only)}
  \addlegendentry{\sys ($\alpha = .4$ in (a), whiskers $\alpha$ .2 to .6)}
  \addlegendimage{sys eagle, marker legend}
  \addlegendentry{EAGLE}
\nextgroupplot[row a]
  \addplot[bar within] coordinates {(A,1.683) (B,1.355) (C,1.281) (D,1.167)};
  \addplot[bar perrow] coordinates {(A,1.720) (B,1.348) (C,1.283) (D,1.167)};
  \addplot[bar hold] coordinates {(A,2.378) (B,1.487) (C,1.371) (D,1.166)};
  \addplot[bar flexit, error bars/.cd, y dir=both, y explicit, error bar style={black, line width=0.6pt}, error mark options={black, mark size=1.2pt, rotate=90}]
    coordinates {(A,2.378) += (0,0.149) -= (0,0.000) (B,1.499) += (0,0.050) -= (0,0.000) (C,1.390) += (0,0.125) -= (0,0.001) (D,1.162) += (0,0.050) -= (0,0.000)};
  \draw[nonee ref] ([xshift=-12pt]axis cs:A,1) -- ([xshift=12pt]axis cs:D,1);
\nextgroupplot[row a]
  \addplot[bar within] coordinates {(A,1.521) (B,1.270) (C,1.216) (D,1.125)};
  \addplot[bar perrow] coordinates {(A,1.939) (B,1.438) (C,1.330) (D,1.184)};
  \addplot[bar hold] coordinates {(A,1.856) (B,1.391) (C,1.210) (D,1.121)};
  \addplot[bar flexit, error bars/.cd, y dir=both, y explicit, error bar style={black, line width=0.6pt}, error mark options={black, mark size=1.2pt, rotate=90}]
    coordinates {(A,1.969) += (0,0.004) -= (0,0.001) (B,1.426) += (0,0.007) -= (0,0.000) (C,1.316) += (0,0.006) -= (0,0.000) (D,1.175) += (0,0.001) -= (0,0.000)};
  \draw[nonee ref] ([xshift=-12pt]axis cs:A,1) -- ([xshift=12pt]axis cs:D,1);
\nextgroupplot[row b, ylabel={Speedup}]
  \draw[nonee ref] (axis cs:-4.3,1) -- (axis cs:1.5,1);
  \addplot[sys within] coordinates {(-0.58,1.121) (-1.85,1.127)};
  \addplot[sys perrow] coordinates {(-0.55,1.12) (-1.86,1.15)};
  \addplot[sys hold] coordinates {(-0.50,1.26) (-1.78,1.50)};
  \addplot[sys flexit] coordinates {(0.00,1.43) (-0.64,1.58) (-1.89,1.59)};
  \addplot[only marks, sys eagle] coordinates {(-0.05,1.45)};
  \node[anchor=south, font=\scriptsize, inner sep=1pt, text=sysflexit] at (axis cs:-0.64,1.63) {1\%};
  \node[anchor=south, font=\scriptsize, inner sep=1pt, text=sysflexit] at (axis cs:-1.89,1.64) {3\%};
  \node[anchor=west, font=\scriptsize, inner sep=1pt] at (axis cs:0.12,1.36) {$\tau{=}\infty$};
\nextgroupplot[row b]
  \draw[nonee ref] (axis cs:-4.3,1) -- (axis cs:1.5,1);
  \addplot[sys within] coordinates {(0.95,1.656) (-1.37,1.563)};
  \addplot[sys perrow] coordinates {(0.98,1.67) (-1.31,1.58)};
  \addplot[sys hold] coordinates {(1.09,1.85) (-1.28,1.79)};
  \addplot[sys flexit] coordinates {(0.07,1.27) (1.01,1.86) (-1.38,1.78)};
  \addplot[only marks, sys eagle] coordinates {(-0.19,1.58)};
  \node[anchor=north west, font=\scriptsize, inner sep=1pt] at (axis cs:0.12,1.25) {$\tau{=}\infty$};
\nextgroupplot[row b]
  \draw[nonee ref] (axis cs:-4.3,1) -- (axis cs:1.5,1);
  \addplot[sys within] coordinates {(-1.34,1.564) (-3.88,1.645)};
  \addplot[sys perrow] coordinates {(-1.42,1.92) (-3.92,2.14)};
  \addplot[sys hold] coordinates {(-1.30,1.71) (-3.80,1.88)};
  \addplot[sys flexit] coordinates {(0.09,0.92) (-1.37,1.91) (-3.90,1.96)};
  \addplot[only marks, sys eagle] coordinates {(0.00,1.17)};
  \node[anchor=east, font=\scriptsize, inner sep=1pt] at (axis cs:-0.08,0.91) {$\tau{=}\infty$};
\end{groupplot}
\node[anchor=north east, font=\scriptsize, fill=white, inner sep=1pt] at ([xshift=-2pt,yshift=-2pt]map c1r1.north east) {(a) 70B, $N = 47$};
\node[anchor=north east, font=\scriptsize, fill=white, inner sep=1pt] at ([xshift=-2pt,yshift=-2pt]map c2r1.north east) {(a) 13B, $N = 80$};
\node[anchor=north east, font=\scriptsize, fill=white, inner sep=1pt] at ([xshift=-2pt,yshift=-2pt]map c3r1.north east) {(a) 8B, $N = 527$};
\node[anchor=south] at ($(map c2r1.north)+(0,0.05cm)$) {\pgfplotslegendfromname{maplegend}};
\node[anchor=north east, font=\scriptsize, align=right, inner sep=0pt] at ([xshift=-3pt,yshift=-8.5pt]map c1r1.south west) {$E/L$\\$p$};
\node[anchor=north east, font=\scriptsize, fill=white, inner sep=1pt] at ([xshift=-2pt,yshift=-2pt]map c1r2.north east) {(b) 70B};
\node[anchor=north east, font=\scriptsize, fill=white, inner sep=1pt] at ([xshift=-2pt,yshift=-2pt]map c2r2.north east) {(b) 13B};
\node[anchor=north east, font=\scriptsize, fill=white, inner sep=1pt] at ([xshift=-2pt,yshift=-2pt]map c3r2.north east) {(b) 8B};
\end{tikzpicture}
\caption{Decode throughput over non-EE (dashed) at full memory. (a)~The synthetic ramp, by exit pattern $(E/L, p)$ with $p \geq E/L$. (b)~CNN/DM on LayerSkip against ROUGE-2: \sys from $\tau = \infty$ to the 1\% and 3\% risk targets, the baselines at both targets, and EAGLE at its best $k$, each over its own target's non-EE.
}
\label{fig:map}
\end{figure*}

At full memory with the synthetic ramp (\cref{fig:map}a), wherever exits pay ($p > E/L$) \sys is the highest-throughput system that forces no exits on 70B and 13B, and on 8B the highest at (.25, .9) and within 1.1\% of the highest elsewhere, while each baseline gains only in its own regime.

\heading{On weight-bound 70B, only holding pays}
Within-iteration batching and per-row skipping gain at most 3.8\% and 7.0\%, because a step with any continuing token still reads every deep weight.
Holding serves several iterations with one read of the deep weights, so \syshold gains up to 2.05$\times$ where $p > E/L$, and speculative holding keeps held requests decoding, so at (.25, .9) and $\alpha = .4$ \sys reaches 2.22--2.24$\times$ the throughput of non-EE on 70B across $N$, 2.1$\times$ the speedup of per-row skipping.

\begin{figure}[t]
\centering
\begin{tikzpicture}[font=\footnotesize]
\begin{groupplot}[
  group style={group size=2 by 1, horizontal sep=0.75cm},
  width=0.5\columnwidth,
  height=2.45cm,
  xmode=log, log basis x=2,
  xmin=0.85, xmax=56,
  xtick={1,2,4,8,16,47},
  xticklabels={1,2,4,8,16,47},
  xlabel={Requests $N$},
  flexit axis,
]
\nextgroupplot[title={$(E/L, p) =$ (.25, .9)}, ymin=1.95, ymax=2.8, ytick={2,2.4,2.8}, ylabel={Speedup},
  legend to name=frontierlegend, legend columns=3, flexit legend,
  legend style={/tikz/every even column/.append style={column sep=0.8ex}}]
  \addplot[sys flexit, only marks] coordinates {
    (1,2.386) (2,2.118) (4,2.300) (8,2.282) (16,2.267) (24,2.320) (32,2.293) (40,2.272) (47,2.254)};
  \addlegendentry{\sys}
  \addplot[sys hold, only marks] coordinates {
    (1,2.189) (2,2.041) (4,2.071) (8,2.025) (16,2.047) (24,2.060) (32,2.041) (40,2.075) (47,2.060)};
  \addlegendentry{\syshold}
  \addplot[black, dashed] coordinates {
    (1,3.075) (2,2.645) (4,2.519) (8,2.445) (16,2.408) (24,2.397) (32,2.386) (40,2.385) (47,2.388)};
  \addlegendentry{best full-flush policy}
  \addplot[black, densely dotted] coordinates {
    (1,3.071) (2,2.547) (4,2.334) (8,2.212) (16,2.149) (24,2.132) (32,2.114) (40,2.116) (47,2.107)};
  \addlegendentry{$S^*_N$}
  \addplot[sysflexit!60, thin, forget plot] coordinates {
    (1,2.327) (2,2.327) (4,2.328) (8,2.331) (16,2.316) (24,2.306) (32,2.309) (40,2.267) (47,2.257)};
  \addlegendimage{black!45, thin}
  \addlegendentry{calibrated model}
  \addplot[syshold!60, thin, forget plot] coordinates {
    (1,2.051) (2,2.053) (4,2.056) (8,2.062) (16,2.074) (24,2.085) (32,2.064) (40,2.086) (47,2.094)};
\nextgroupplot[title={(.25, .5)}, ymin=1.1, ymax=1.65]
  \addplot[sys flexit, only marks] coordinates {
    (1,1.445) (2,1.425) (4,1.407) (8,1.427) (16,1.396) (24,1.399) (32,1.387) (40,1.400) (47,1.396)};
  \addplot[sys hold, only marks] coordinates {
    (1,1.196) (2,1.180) (4,1.177) (8,1.176) (16,1.180) (24,1.208) (32,1.203) (40,1.208) (47,1.199)};
  \addplot[black, dashed] coordinates {
    (1,1.599) (2,1.559) (4,1.510) (8,1.486) (16,1.478) (24,1.470) (32,1.470) (40,1.466) (47,1.464)};
  \addplot[black, densely dotted] coordinates {
    (1,1.599) (2,1.411) (4,1.292) (8,1.247) (16,1.226) (24,1.215) (32,1.213) (40,1.208) (47,1.206)};
  \addplot[sysflexit!60, thin] coordinates {
    (1,1.441) (2,1.441) (4,1.441) (8,1.441) (16,1.421) (24,1.440) (32,1.439) (40,1.439) (47,1.438)};
  \addplot[syshold!60, thin] coordinates {
    (1,1.187) (2,1.188) (4,1.189) (8,1.192) (16,1.197) (24,1.202) (32,1.207) (40,1.212) (47,1.216)};
\end{groupplot}
\node[anchor=south] at ($(group c1r1.north)!0.5!(group c2r1.north)+(0,0.42cm)$) {\pgfplotslegendfromname{frontierlegend}};
\end{tikzpicture}
\caption{Llama-2-70B at $\alpha = .4$ after the controller's warm-up against the optima in W (\cref{thm:hold,thm:spec}): the best full-flush policy and the holding optimum $S^*_N$, with the calibrated model (thin lines).}
\label{fig:frontier}
\end{figure}

\heading{Near the optimum, and past the best policy without speculation}
On 70B at $\alpha = .4$ and $N \geq 16$, \sys reaches 92--96\% of the best full-flush policy in W and \syshold 94--99\% of the holding optimum $S^*_N$, although the host sees exits two steps late, and \sys beats $S^*_N$ from $N = 8$ and 2 at (.25, .9) and (.25, .5) (\cref{fig:frontier}).
The controller, never told $\alpha$, is within 2.5\% of the best fixed $(k, m)$ on 70B and 13B and 8.0\% on 8B, and plan-ahead flushing is worth 8--11\% on 70B and 34--69\% on 13B and 8B over a host sync every step (\cref{sec:appendix-overhead}).

\heading{With tensor parallelism, 70B moves toward the ridge, and holding still pays}
Sharded over two GPUs, 70B holds 280 requests at full memory instead of 47, so skipped rows begin to save compute: per-row skipping rises from 0.98$\times$ at $N = 16$ to 1.63$\times$ at full memory.
\sys still reaches 2.12--2.35$\times$ the throughput of non-EE across $N$, 1.44$\times$ the speedup of per-row skipping at full memory, with its controller on the single-GPU calibration (\cref{tab:tp}).

\heading{On attention-bound 13B, holding goes beyond per-token savings}
Exits also skip the deep attention that dominates a 13B step, so per-row skipping gains 1.72$\times$ at (.25, .9) and holding 2.38$\times$, 38\% more.
Tentative rows pay the attention that held rows skip, so at $\alpha \leq .4$ and $N \geq 64$ the controller holds on 83--100\% of steps (\cref{sec:appendix-grid}).

\heading{On compute-bound 8B, compaction pays}
With 527 requests, rows above the ridge cost time, so the systems that compact the deep pass win: \sys beats the better in-place policy by 2.5--8.8\% where $p \geq E/L$, holding on 95--99\% of steps even at $\alpha = .4$, so \sys runs 0.99--1.02$\times$ per-row skipping, the best policy past the ridge, and 1.97$\times$ non-EE at (.25, .9) (\cref{tab:map8}).

\heading{Do no harm}
Wherever $p \leq E/L$, including where no token exits, the deployed system stays within 1.8\% of non-EE on 13B and 70B, and within 4.2\% on 8B, where the compacted pass would cost 6--10\%, so a deployment whose model rarely exits runs in place (\cref{sec:appendix-grid}).
In five of these cells, four with no exits, it gets there through the fallback ($\ddagger$), and compact mode, which has no fallback, stays within 1.8\% on 13B and 70B.

\subsection{Faithful and Efficient on a Real EE Model}
\label{sec:eval-layerskip}

The softmax confidence ramp follows layer 20 of 80 on 70B, layer 8 of 32 on 8B, and, on 13B, layer 15 of 40 for summarization and layer 10 for PG-19, with a threshold per exit layer, task, and target, for an exit rate $p^* = .5$ or a risk target $\epsilon^* \in \{1\%, 3\%\}$ (\cref{sec:appendix-layerskip}).
A risk target is the rate at which exited tokens disagree with the non-EE model in teacher-forced greedy calibration, so it bounds neither the task's loss nor disagreement under sampling.
Summarization is greedy, and PG-19~\cite{pg19} runs sample at top-p .95 and temperature 1 under coupled guessing.

\begin{figure}[t]
\centering
\begin{tikzpicture}[font=\footnotesize]
\begin{groupplot}[
  group style={group size=2 by 1, horizontal sep=0.3cm, y descriptions at=edge left},
  scale only axis,
  width=0.33\columnwidth,
  height=2.5cm,
  y dir=reverse,
  ymin=0.5, ymax=7.5,
  xmin=-36, xmax=11, xtick={-30,-15,0},
  ytick={1,2,3,4,5,6,7},
  yticklabels={GSM8K 13B, BBH 13B, MBPP 13B, HumanEval 13B, CNN/DM 13B, GSM8K 70B, GSM8K 8B},
  y tick label style={font=\scriptsize},
  x tick label style={font=\scriptsize},
  flexit axis, ymajorgrids=false, xmajorgrids=true,
  title style={font=\scriptsize, yshift=-2pt},
  every axis plot/.append style={only marks, line width=0.5pt},
]
\nextgroupplot[title={EE model, B $-$ A}, legend to name=qlegend, legend columns=3, flexit legend]
  \draw[black!55, line width=0.5pt] (axis cs:0,0.5) -- (axis cs:0,7.5);
  \addplot[color=black!55, mark=*, mark size=1.3pt] coordinates {(-10.0,0.78) (-10.8,1.78) (-8.4,2.78) (-1.2,3.78) (0.7,4.78) (-14.0,5.78) (-21.2,6.78)};
  \addlegendentry{$\epsilon^*$ 1\%}
  \addplot[color=black!55, mark=square*, mark size=1.2pt] coordinates {(-17.6,1.00) (-17.8,2.00) (-11.2,3.00) (-6.1,4.00) (-0.4,5.00) (-29.0,6.00) (-32.8,7.00)};
  \addlegendentry{$\epsilon^*$ 3\%}
  \addplot[color=black!55, mark=triangle*, mark size=1.5pt] coordinates {(-4.8,1.22) (-3.8,2.22) (-12.0,3.22) (1.2,4.22) (0.5,5.22) (-18.0,6.22) (-21.2,7.22)};
  \addlegendentry{$p^*$ .5}
\nextgroupplot[title={\sys $-$ B, 95\% CI}]
  \draw[black!55, line width=0.5pt] (axis cs:0,0.5) -- (axis cs:0,7.5);
  \addplot[sys flexit, mark=*, mark size=1.3pt, error bars/.cd, x dir=both, x explicit, error bar style={line width=0.5pt}, error mark=none]
    coordinates {(1.2,0.78) += (2.4,0) -= (2.4,0) (-0.2,1.78) += (1.2,0) -= (1.2,0) (0.0,2.78) += (1.6,0) -= (1.6,0) (0.0,3.78) += (0.0,0) -= (0.0,0) (0.3,4.78) += (0.5,0) -= (0.3,0) (4.0,5.78) += (6.0,0) -= (5.0,0) (-1.2,6.78) += (2.8,0) -= (2.8,0)};
  \addplot[sys flexit, mark=square*, mark size=1.2pt, error bars/.cd, x dir=both, x explicit, error bar style={line width=0.5pt}, error mark=none]
    coordinates {(-1.2,1.00) += (1.2,0) -= (1.6,0) (-0.4,2.00) += (1.4,0) -= (1.6,0) (0.4,3.00) += (1.2,0) -= (1.2,0) (0.0,4.00) += (0.0,0) -= (0.0,0) (-0.4,5.00) += (0.4,0) -= (0.5,0) (2.0,6.00) += (4.0,0) -= (4.0,0) (-1.2,7.00) += (2.4,0) -= (2.8,0)};
  \addplot[sys flexit, mark=triangle*, mark size=1.5pt, error bars/.cd, x dir=both, x explicit, error bar style={line width=0.5pt}, error mark=none]
    coordinates {(-0.8,1.22) += (2.0,0) -= (2.0,0) (1.0,2.22) += (1.6,0) -= (1.4,0) (-0.4,3.22) += (1.6,0) -= (1.6,0) (0.0,4.22) += (3.7,0) -= (3.7,0) (-0.1,5.22) += (0.4,0) -= (0.6,0) (1.0,6.22) += (7.0,0) -= (7.0,0) (-0.8,7.22) += (2.8,0) -= (2.8,0)};
\end{groupplot}
\node[anchor=south] at ($(group c1r1.north)!0.5!(group c2r1.north)+(0,0.4cm)$) {\pgfplotslegendfromname{qlegend}};
\node[anchor=north, font=\scriptsize] at ($(group c1r1.south)!0.5!(group c2r1.south)+(0,-0.45cm)$) {points (accuracy or ROUGE-2)};
\end{tikzpicture}
\caption{Quality at $N = 64$ on LayerSkip: the EE model's own loss (B $-$ A, B = batch-1 EE, A = the non-EE model) and \sys $-$ B at three targets per task, at one exit layer per model (13B: layer 15, 70B: 20, 8B: 8). \Cref{tab:quality} adds a second exit layer.
}
\label{fig:quality}
\end{figure}

\heading{\sys adds no statistically significant loss of its own}
On held-out GSM8K~\cite{gsm8k}, BBH~\cite{bbh}, MBPP~\cite{mbpp}, HumanEval~\cite{chen2021humaneval}, and CNN/DM~\cite{cnndm}, the interval of \sys minus batch-1 EE contains 0 in 40 of 42 cells and favours \sys in the other two (\cref{fig:quality}), so the experiment resolves no degradation from batching, although it does not prove equivalence.
The EE model's own loss (B $-$ A) reaches 33 points on GSM8K and none on CNN/DM, so we claim faithfulness to the EE model, not losslessness.

\heading{Fidelity}
Batching changes none of \sys's exit decisions beyond floating-point near-ties: on identical prefixes, its decisions at $N = 64$ and $N = 1$ agree on 99.71--99.88\% of tokens, each flip within 0.042 of the threshold, and under vLLM's batch-invariant kernels all 20,055 decisions on 13B agree, under every policy.
Whole documents still differ more often than non-EE's (68--70\% against 84--88\% identical), mostly from a full-depth token (\cref{sec:appendix-layerskip}).

\heading{\sys gains most where the EE model exits most}
On CNN/DM summarization task at full memory (\cref{fig:map}b), \sys reaches 1.58--1.91$\times$ the throughput of non-EE at $\epsilon^*$ 1\% and $-$1.4 to +1.0 ROUGE-2 points, and 12\% and 40\% higher throughput than per-row skipping on 13B and 70B and level with it on 8B.
The regimes carry over: speculation lifts weight-bound 70B from 1.26$\times$ for \syshold to 1.58$\times$, the two variants match on 13B, and compaction pays on compute-bound 8B (\cref{sec:appendix-sd}).
On sampled PG-19, with fewer exits, \sys runs 1.11--1.49$\times$ at the risk targets (8B: 0.96--1.28$\times$, \cref{tab:layerskip-speed}).

\subsection{Early Exit or Speculative Decoding?}
\label{sec:eval-sd}

\begin{figure}[t]
\centering
\begin{tikzpicture}[font=\footnotesize]
\begin{groupplot}[
  group style={group size=2 by 1, horizontal sep=1.1cm},
  scale only axis,
  width=0.355\columnwidth,
  height=2.05cm,
  xmode=log, log basis x=2,
  xmin=13, xmax=640,
  xtick={16,128,528}, xticklabels={16,128,528}, extra x ticks={64,256,384}, extra x tick labels={},
  xlabel={Requests $N$},
  flexit axis,
]
\nextgroupplot[ylabel={Speedup}, ymin=0.9, ymax=1.9, ytick={1,1.25,1.5,1.75},
  ylabel style={yshift=-3pt},
  legend to name=xlegend, legend columns=3, flexit legend]
  \draw[nonee ref] (axis cs:13,1) -- (axis cs:640,1);
  \node[anchor=north west, font=\scriptsize\bfseries, inner sep=2pt] at (rel axis cs:0,1) {(a)};
  \addplot[sys flexit] coordinates
    {(16,1.183) (64,1.280) (128,1.470) (256,1.640) (384,1.765) (528,1.787)};
  \addlegendentry{\sys (.25, .9)}
  \addplot[sys flexit, color=sysflexit!60, dashed, mark=o, mark options={solid}] coordinates
    {(16,1.028) (64,1.048) (128,1.061) (256,1.136) (384,1.198) (528,1.235)};
  \addlegendentry{(.25, .5)}
  \addplot[sys flexit, color=sysflexit!40, densely dotted, mark=diamond, mark size=1.8pt, mark options={solid}] coordinates
    {(16,0.935) (64,0.952) (128,1.021) (256,1.065) (384,1.095) (528,1.117)};
  \addlegendentry{(.5, .5)}
  \addplot[sys flexit, dashdotted, mark=pentagon*, mark size=1.6pt, mark options={solid}] coordinates
    {(16,1.019) (64,1.090) (128,1.113) (256,1.149) (384,1.208) (527,1.275)};
  \addlegendentry{LayerSkip-8B, 3\%}
  \addplot[sys eagle] coordinates
    {(16,1.249) (64,1.165) (128,1.167) (256,1.117) (384,1.068) (528,0.989)};
  \addlegendentry{EAGLE-3}
  \addplot[sys perrow] coordinates
    {(16,0.922) (64,1.076) (128,1.265) (256,1.452) (384,1.591) (528,1.698)};
  \addlegendentry{Per-row skip.\ (.25, .9)}
\nextgroupplot[ylabel={$\beta^*$}, ymin=0, ymax=1, ytick={0,0.5,1}, ylabel style={yshift=-6pt}]
  \fill[syseagle!25] (axis cs:13,0.351) rectangle (axis cs:640,0.380);
  \node[anchor=north west, font=\scriptsize\bfseries, inner sep=2pt] at (rel axis cs:0,1) {(b)};
  \node[anchor=south west, font=\scriptsize, inner sep=1pt] at (axis cs:14,0.380) {EAGLE-3};
  \addplot[sys flexit] coordinates
    {(16,0.275) (64,0.348) (128,0.671) (256,0.908) (384,0.946) (528,0.954)};
  \addplot[sys flexit, color=sysflexit!60, dashed, mark=o, mark options={solid}] coordinates
    {(16,0.108) (64,0.125) (128,0.206) (256,0.572) (384,0.673) (528,0.717)};
  \addplot[sys flexit, color=sysflexit!40, densely dotted, mark=diamond, mark size=1.8pt, mark options={solid}] coordinates
    {(16,0.000) (64,0.000) (128,0.161) (256,0.475) (384,0.530) (528,0.567)};
\end{groupplot}
\node[anchor=south] at ($(group c1r1.north)!0.5!(group c2r1.north)+(0,0.05cm)$) {\pgfplotslegendfromname{xlegend}};
\end{tikzpicture}
\caption{Llama-3.1-8B, sampled PG-19. (a)~EAGLE-3's speedup falls with $N$ while \sys's grows (synthetic ramp at $\alpha = .4$, LayerSkip-8B at $\epsilon^*$ 3\%). (b)~The break-even acceptance $\beta^*$ a draft head needs to match \sys rises past EAGLE-3's acceptance (purple band).
}
\label{fig:crossover}
\end{figure}

\heading{On today's EE models, the task decides}
EAGLE's heads serve chat models, so we compare speedups, each over its own target's non-EE run.
On CNN/DM summarization task at full memory, \sys at $\epsilon^*$ 1\% speeds up its target 9--63\% more than EAGLE does at its best $k$ on every model, most on compute-bound 8B (1.91 against 1.17$\times$, \cref{fig:map}b).
EAGLE leads at $N = 16$, and \sys overtakes it from $N = 64$ on 13B and 70B and at full memory on 8B (\cref{tab:summ}).
With $\tau = \infty$, \sys exits nothing and verifies every guess.
This is LayerSkip's self-speculation~\cite{elhoushi2024layerskip}, lossless up to floating-point near-ties, and it matches EAGLE on 70B CNN/DM from $N = 64$.
On sampled PG-19, \sys beats this self-speculation by 5--21\% at $\epsilon^*$ 1\% and 16--54\% at 3\% (\cref{sec:appendix-layerskip}).
Trusting exits is a quality choice, which relaxed acceptance~\cite{narasimhan2025cascades} offers draft heads too.
On PG-19, EAGLE beats \sys on 13B and 70B by 7--17\% at $\epsilon^*$ 3\% and 24--36\% at 1\% (\cref{tab:eagle-pg19}).

\heading{With an EE model at (.25, .9), large batches favour early exit}
EAGLE-3's speedup on Llama-3.1-8B falls from 1.25$\times$ at $N = 16$ to 0.99$\times$ at full memory with sampled text, and from 2.26$\times$ to 1.32$\times$ with greedy decoding, although its acceptance does not change with $N$ (\cref{fig:crossover}, greedy in \cref{tab:crossover}).
A verify step runs the full model over $N(k+1)$ rows, which cost time in proportion past the ridge, so the best $k$ falls to 1 or 2 and drafting stops paying, where an adaptive scheme~\cite{su2023synergy,nightjar2025,adaspec2025} would turn it off (1.00$\times$).
Trusted exits remove deep rows instead, so \sys's speedup in its deployed compact mode grows from 1.18$\times$ to 1.79$\times$ sampled and from 1.30$\times$ to 1.93$\times$ greedy, overtaking EAGLE-3 from $N = 64$ sampled and 256 greedy.
Per-row skipping trails \sys at every sampled $N$ and overtakes EAGLE-3 only from $N = 128$, so trusting exits wins at large batches, and holding moves the crossover earlier.

\heading{A break-even acceptance explains the winner}
We extend the calibrated cost model with a draft head (\cref{sec:appendix-sd}), and at each cell's measured tokens per verify it predicts EAGLE's speed within a median of 0.94--1.04 per model.
At \sys's measured speed, the break-even acceptance $\beta^*$ rises from .28 at $N = 16$ to .95 at full memory at (.25, .9) (\cref{fig:crossover}b), and comparing it with EAGLE-3's measured acceptance in the same cell picks the higher-throughput system in 34 of 36 cells, greedy and sampled.

\section{Discussion and Limitations}
\label{sec:discussion}

Today's EE models are the main limitation: none we serve reaches the exit patterns where \sys gains most without a large quality loss, so these patterns are a target for EE models, such as those trained for early exits~\cite{varshney2024lite,pan2024eetuning,jamialahmadi2025balcony}.
Exit potential shrinks across model generations but is highest in dense models and above 20B parameters~\cite{eediminish2026}, as on 70B here.
Held and tentative tokens wait for their flush, so \sys cuts the TPOT but raises the median p99 largest gap on 70B to 2.64$\times$ non-EE's, and under a gap bound of 1.5$\times$ non-EE's TPOT it reverts to holding, at 1.48--1.54$\times$ non-EE (\cref{sec:appendix-tails}).
Our claims cover dense models, where a flush reads each deep weight once, one ramp, and a single guess.
Pipeline parallelism would idle the deep stages between flushes.
Compact mode refuses penalties, log-probabilities, and structured output.
Synthetic exits and guesses are independent, yet live LayerSkip runs within 5\% of its synthetic twin in 49 of 54 cases (\cref{sec:appendix-layerskip}).

\section{Related Work}
\label{sec:related}

Grouped policies~\cite{miao2024efficient} and SkipDecode~\cite{skipdecode} override tokens' decisions, Apparate~\cite{apparate2024} runs exits through the deep layers, FREE~\cite{bae2023free}, AdaDecode~\cite{adadecode2025}, and EE-LLM~\cite{eellm} run an exit's skipped layers with its sequence's next token, HELIOS~\cite{kumar2025helios} switches among EE models, E3~\cite{e3} rebatches the inputs that continue past an exit so that later blocks keep full batches, and concurrent vSkipper~\cite{vskipper2026} is per-row skipping.
None holds many requests' continuing tokens at the exit layer so that one read of the deep weights serves several steps, and continuous depth-wise batching~\cite{mor2025,cdb2026} groups looped models' tokens by loop step on shared weights, not on distinct layers.
Speculative decoding~\cite{speculative2023,eagle3}, self-speculation~\cite{elhoushi2024layerskip,speed2023,zhang2024draftverify,xia2025swift,kangaroo}, its hybrids with EE~\cite{xu2025specee,liu2024eesd,hispec2025}, and batch-aware variants~\cite{magicdec2025} verify every draft with the full model, and BiLD~\cite{kim2023bild} and speculative cascades~\cite{narasimhan2025cascades} trust confident drafts but still run the large model over each, while \sys's trusted tokens skip the deep layers.

\section{Conclusion}
\label{sec:conclusion}

We introduced \sys, a serving framework that turns early exits into throughput through cross-iteration batching.
Continuing tokens share a deep-layer pass across decoding steps, amortizing weight reads.
\sys chooses from two policies: Holding pauses requests until that pass, and speculative holding advances them using exit-layer guesses verified at the next pass.
A controller selects between the two policies, while compacted execution and plan-ahead flushing make asynchronous scheduling compatible.

On LayerSkip models summarizing the CNN/DM dataset, \sys achieves 1.58--1.91$\times$ speedup within 1.4 ROUGE-2 points of non-EE.
\sys never forces exits and adds no quality loss relative to batch-1 EE, providing a practical foundation for serving increasingly capable EE models.

\clearpage
{
\small
\bibliographystyle{plain}
\bibliography{ref}
}

\clearpage
\newpage
\appendix

\setcounter{topnumber}{4}
\setcounter{dbltopnumber}{4}
\renewcommand{\topfraction}{0.95}
\renewcommand{\dbltopfraction}{0.95}
\renewcommand{\textfraction}{0.05}

\section{Proofs and Dynamic Programs}
\label{sec:appendix-proofs}

The model W, the faithful class, and the notation are those of \cref{sec:theory-model}.
A position's token is undetermined until its ramp decides and, if it continues, its deep pass runs.
Policies observe exits and the depths that follow from them, but not guess outcomes before the flush, nor any signal correlated with them, except for the guess-outcome oracle of \cref{tab:spec-optimum}.
W charges a step $|\Lambda| \cdot t_w$ whatever its rows, including the steps the proofs construct, so added rows, tentative and verification rows included, are free.
Proofs that compare two policies couple them: each request's position gets one draw of its exit, guess outcome and confidence from their joint law, which both policies use, and in W the draws are i.i.d.\ across positions and independent of the tokens.

\heading{Lemma 1 (whole sweeps suffice)}
\emph{For any non-speculative faithful policy there is one at least as fast in which every step reads the shallow sweep $1..E$, the deep sweep $E{+}1..L$, or both, each over some set of rows.}
Given $\pi$, let $\pi'$ run $\pi$ as a simulation, start each position when $\pi$ does, and change only which rows each step reads, by two rules.
\emph{Lazy reads}: it postpones every read of a layer $\ell \notin \{E, L\}$, which reveals nothing, and performs the postponed reads of $\ell$, merged, only in a step that reads $\ell+1$ over a row that needs them.
\emph{Eager inclusion}: every read it performs covers every row whose previous layer is done, with the exited positions whose deep KV that row needs.
So each step of $\pi'$ reads $a..E$ and $b..L$, and a row in a read of $a > 1$ finished $a-1$ in an earlier step that also read $a..E$, where eager inclusion took it through $E$, so $a = 1$ and likewise $b = E+1$.
Each read of $\ell$ by $\pi'$ takes over a read of $\ell$ by $\pi$ from the same or an earlier step that no other read took over, so $\pi'$ never trails $\pi$ in layer reads and determines every position no later.
It hands each exit to the simulated $\pi$ when $\pi$ would see it, so it is online, and it computes the same tokens, so it is faithful.
\Cref{thm:spec} needs no speculative version of this lemma, because full-flush policies (\cref{sec:theory-spec}) read whole sweeps by definition, and whether \cref{thm:spec} extends to every single-guess policy is open.

\heading{Lemma 2 (rows are free)}
\emph{Below the ridge, (1) an involuntary stay never shortens a critical path, (2) idling a subset of ready requests never helps, and only skipping the whole shallow sweep in a step (a deep-only flush) can, and (3) pausing a request never helps a single-guess speculative policy, even with a pause rule that depends on exit-layer confidence.}
Each part builds $\pi'$ from $\pi$ by adding rows to the same reads, simulating $\pi$ as in Lemma 1.
(1) $\pi'$ lets $\pi$'s forced stays exit, which determines them at $E$ instead of $L$, and under the coupling the later positions keep their outcomes even when the token changes, so by induction every position is determined no later.
(2) $\pi'$ adds an idled ready row to the shallow sweep that $\pi$ runs anyway, and by eager inclusion to later reads, and the same induction applies.
(3) $\pi'$ copies $\pi$'s steps and flushes but decodes every request in every shallow step.
Positions on batch-1 EE's trajectory share their draws, and positions after an unresolved wrong guess, which are rolled back, get their own, which $\pi'$ samples for the simulated $\pi$.
Let $m$ be a request's committed index at a flush and $n$ the positions it decoded since the previous flush. After the flush its committed index is $f(m, n) = \min(m + n, j^*(m))$, where $j^*(m)$ is the first continuing position after $m$ whose guess is wrong.
If $m' \geq m$, then $n' \geq n$ gives $m' + n' \geq m + n$, and the shared draws give $j^*(m') \geq j^*(m)$ (equal if $j^*(m) > m'$), so $f(m', n') \geq f(m, n)$, and by induction over flushes $\pi'$ commits at least as many tokens in the same time.
The argument needs no independence between guesses and confidence, since $\pi'$ ignores confidence, and it fails where rows cost time (\cref{sec:theory-beyond}).

\heading{Proof of \cref{thm:batch1}}
A request's positions run in sequence, because each needs its predecessor's token, and a position starts no earlier than the step after its predecessor's token is determined.
An exiting position reads at least $E$ layers and a continuing one at least $L$, each read costs $t_w$ of wall time in its step, and steps that read layers for other rows only add time.
Over a horizon $T$, each request $a$ needs $T \geq t_w (E X_a + L Y_a)$, where $X_a$ and $Y_a$ count its exiting and continuing positions determined by $T$, and summing over the $N$ requests bounds the speedup by $L / \big(p_T E + (1 - p_T) L\big)$, where $p_T$ is the exit fraction of those tokens, which tends to $p$.

\heading{Fixed periods are below $\Gone$}
With $f(k) = 1 - k + \sum_{j=1}^{k} p^j$, $S(k) < \Gone$ is equivalent to $r f(k) < p^k$, which holds because $f(k) - p^k \leq 0$ and $r < 1$, for every $k \geq 1$, $0 < p < 1$, and $0 < r < 1$.

\heading{Proof of \cref{thm:hold} and the computed margins}
By Lemmas 1 and 2 and i.i.d.\ exits, the state is the number $n$ of ready requests at the start of a step, and there are three actions: a shallow step (cost $E$, reward the ready count, continuers become held, next state binomial with parameters $n$ and $p$), a flush step (cost $L$, the shallow sweep for ready rows plus the deep sweep for held rows and this step's continuers, reward the ready count, next state $N$), and a deep-only flush (cost $L - E$, reward 0, next state $N$).
Cycles from $N$ back to $N$ form a renewal process, so $\lambda^*$ is the root of $h(\lambda) = W_\lambda(N)$, the best expected cycle reward minus $\lambda$ times its cost~\cite{dinkelbach1967}, where $W_\lambda(0) = -\lambda (L - E)$ and, with $b_{n,j}$ the binomial probabilities and $A_n = n - \lambda E + \sum_{j<n} b_{n,j} W_\lambda(j)$,
\[
W_\lambda(n) = \max\big\{ A_n / (1 - p^n),\; n - \lambda L,\; -\lambda (L - E) \big\}.
\]
The shallow term solves its self-loop of probability $p^n$, and $h$ strictly decreases, so bisection on $[0, N/E]$ finds $\lambda^*$, and $S^*_N = \lambda^* L / N$.
Within a cycle the ready count never increases, since exits stay ready and continuers are held until the flush, and a flush step costs exactly a shallow step plus a deep-only flush, so with the rate fixed at $\lambda^*$ a further shallow step pays iff it yields at least $\lambda^* E$ tokens.
This is a monotone stopping problem, for which the one-step-lookahead rule is optimal: run a deep-only flush once fewer than $\lambda^* E$ requests are ready.
\Cref{tab:optimum} lists $S^*_N$ and the fixed period's share of it, and an independent re-implementation reproduces it.

\begin{table}[t]
\caption{The holding optimum $S^*_N$ in W against the batch-1 gain and a fixed flush period.}
\label{tab:optimum}
\centering
\scriptsize
\setlength{\tabcolsep}{2pt}
\begin{tabular}{@{}l cc cccccc cc@{}}
\toprule
 & & & \multicolumn{6}{c}{$S^*_N$ at $N =$} & \multicolumn{2}{c}{fixed / $S^*_N$} \\
\cmidrule(lr){4-9}\cmidrule(lr){10-11}
$(E/L, p)$ & $\Gone$ & $\max_k S$ & 1 & 4 & 16 & 32 & 64 & 256 & 16 & 64 \\
\midrule
(.25, .9) & 3.08 & 2.09 & 3.08 & 2.33 & 2.15 & 2.12 & 2.10 & 2.09 & .972 & .993 \\
(.375, .7) & 1.78 & 1.25 & 1.78 & 1.35 & 1.28 & 1.27 & 1.26 & 1.25 & .975 & .992 \\
(.25, .5) & 1.60 & 1.20 & 1.60 & 1.30 & 1.22 & 1.21 & 1.20 & 1.20 & .983 & .996 \\
(.5, .5) & 1.33 & 1.00 & 1.33 & 1.09 & 1.04 & 1.03 & 1.02 & 1.01 & .960 & .980 \\
\bottomrule
\end{tabular}
\end{table}

\heading{Proof of \cref{thm:acceptance}}
Tentative work survives iff every earlier guess in its window is right, and with one guess per position nothing else can validate it.
Exits are trusted, so they are accepted with probability 1, and a position after $c$ unresolved continuing positions commits iff all $c$ guesses are correct, an event the policy cannot influence because its decisions ignore guess outcomes.
With i.i.d.\ guesses the probability is $\alpha^c$, and if a request decodes in every step, the number of continuing positions before position $s$ of its window, which counts from 0, is binomial with parameters $s$ and $1-p$, and $\mathbb{E}[\alpha^{c_s}] = (p + (1-p)\alpha)^s$.
For the remark, $\peff = P(\text{exit}) + P(\text{continue and agree}) = P(\text{agree}) + P(\text{exit and disagree}) = (1 - e) + w$.

\heading{Proof of \cref{thm:spec} and the computed margins}
Every deep read of a full-flush policy resolves every unresolved position, and by Lemma 2(3) pausing never helps, so every request that can advance decodes in every shallow step.
The state is the histogram $(n_0, \dots, n_C)$ of requests by unresolved continuing positions, a shallow step moves each request from depth $c$ to $c+1$ with probability $1-p$, and its expected reward is the yield $Y = \sum_c n_c \alpha^c$.
Replacing realized rewards by $\alpha^c$ is exact, because guess outcomes are revealed only at the flush, after which every request restarts at depth 0, and a simulation with real rollbacks agrees within 0.05\%.
(a)~By linearity, a fixed window of $k$ steps achieves exactly $S(k; \peff, r)$ at every $N$ with $Nk \leq \theta^*$.
(b)~Within a window the yield never increases, because depths only grow, and in W a flush step costs exactly a shallow step plus a deep-only flush, so splitting it lets the policy decide after seeing the exits.
This is a monotone stopping problem, for which the one-step-lookahead rule, flush once $Y < \lambda^*_\alpha E$, is optimal.
The rule needs no foresight, so revealing future exits does not change the optimum, and the exact policy at $\lambda^*_\alpha$ equals the rule in every state we enumerated (up to 125,970).
\Cref{tab:spec-optimum} gives the fixed window's share of the best full-flush policy.
With guesses, $\Gone$ is a reference, not a ceiling (1.83 $> \Gone = 1.60$ for one request at (.25, .5), $\alpha = .6$).
We solve this Markov decision process exactly for $N \leq 8$ with a depth cap of 12 (caps of 12 and 20 agree to $10^{-6}$), and use the perfect-information value, which equals the optimum by (b), for larger $N$, with a horizon of $\max(64, 16 k^*)$ steps.

\begin{table}[t]
\caption{A fixed window's share of the best full-flush policy in W (\cref{sec:theory-spec}).}
\label{tab:spec-optimum}
\centering
\scriptsize
\setlength{\tabcolsep}{3pt}
\begin{tabular}{@{}l ccc@{}}
\toprule
Policy information & $N = 16$ & 32 & 64 \\
\midrule
Exits observed, $\alpha \geq .2$ & .978--.995 & .989--.998 & .994--.999 \\
Guess outcomes revealed, $\alpha \geq .2$ & $\geq$ .972 & $\geq$ .985 & $\geq$ .994 \\
Holding only ($\alpha = 0$) & .960--.984 & .972--.991 & .980--.996 \\
\bottomrule
\end{tabular}
\end{table}

\heading{Coupled guessing (\cref{sec:design-spec})}
vLLM samples $y = \arg\max(\ell + G)$ with Gumbel noise $G$ keyed by the request seed and the position, after each side's own top-k and top-p truncation.
\sys guesses $g = \arg\max(\ell_E + G)$ with the deep token's own $G$ and accepts iff $g = y$, so the emitted token is always $y$ and outputs equal batch-1 EE per seed, up to floating-point near-ties.
This requires that the guess uses the target's noise rather than a draft salt, that the fused sampler is off, that history-dependent logits processors apply per position with that position's prefix, and that the ramp does not read the noise.
Every system in \cref{sec:eval}, non-EE and EAGLE included, runs with the fused sampler off and seeds each request by its index, so all arms draw the same noise.

\section{Cost Model and Calibration}
\label{sec:appendix-costmodel}

\heading{Step model}
The controller predicts throughput with a closed-system model of one window of $k$ steps.
A step's GPU time sums, over the layers it runs, the GEMM time $\max\big(W_\ell / (B \eta_g),\, 2 P_\ell r / (F \mu)\big)$ for $r$ rows, $W_\ell$ weight bytes and $P_\ell$ parameters, decode attention over the active rows' contexts, KV writes, the LM head and sampler, and the riding rows' K and V projections, plus fixed per-layer and per-step costs.
$B$ is the memory bandwidth, $F$ the dense BF16 peak, $\mu$ the GEMM efficiency at the compute roof, and $\eta_g$, $\eta_a$ the achieved bandwidth fractions of GEMMs and attention, so the ridge is $\theta^* = F\mu / (B \eta_g)$.
A step costs the maximum of its GPU time and its class's host time, a line in $N$, because asynchronous scheduling hides host work only while it is shorter.
The compact terms price the deep bucket's rows, duplicated block-table rows (a reuse factor), and the riding rows in the QKV GEMMs.

\heading{Calibration}
Three calibrations fit the constants, all on runs that predate the results of this paper, so every measurement in \cref{sec:eval} is out of sample in time: it comes from runs that no fit saw, though the earlier runs cover the same models and many of the same $N$ and exit patterns.
The first fits the in-place constants on non-EE steps and one flush constant on within-iteration batching without exits, so no configuration with exits enters the fit ($\eta_g = 0.703$, $\eta_a = 0.881$, $\theta^* = 311$ rows).
The second adds the compact terms, fitted on 84 points of our first compact runs, and the third refits them on 155 points from those runs and an earlier $N$ sweep, with 471 points held out, among them an earlier run of every synthetic cell (rms error 1.6\% on the fitted points and 2.4\% on the held-out ones): a fixed cost of 16.0\,$\mu$s per layer for multi-head and 10.2\,$\mu$s for grouped-query attention, a small-batch credit on 70B, 0.105\,ms of host time per compact step, a reuse factor of 0.315, and $\mu = 0.496$, which puts the compact engine's ridge at 200--221 rows.
The controller runs the second calibration.
On 8B, whose steps are short enough that host time and sampling decide many of them, the controller adds three terms: the sampler's 5.5\,$\mu$s per sampled row (fitted on runs that predate the results), a scale on the host lines' per-row slope from its own steady steps of at least 256 rows, and the batch's running mean context instead of its starting one, so that $k$ follows the context as a static batch decodes.

\heading{Accuracy}
Over the synthetic cells and the $N$ sweep after the controller's warm-up, measured over predicted speedup (third calibration) has a median of 0.981--1.000 per arm, and every cell at $N \geq 16$ is within $\pm$5\% except four 13B cells at 0.941--0.948.

\section{Additional Synthetic Results}
\label{sec:appendix-grid}

\Cref{tab:map} gives every value of \cref{fig:map}a, the exit patterns it leaves out, and $N = 16$ for 70B and 13B, \cref{tab:map8} for 8B, \cref{tab:map-base} the two baselines, and \cref{tab:tp} 70B with tensor parallelism 2.
With tensor parallelism 2, every system runs on two GPUs of one Grace NUMA node, pinned to the same KV blocks, where 280 requests fill memory, and the controllers keep their single-GPU calibration.
All 40 of its cells are valid.

\begin{table}[t]
\caption{Llama-2-70B with tensor parallelism 2 at (.25, .9): decode throughput over non-EE at the same $N$ (280: full memory), and \sys over per-row skipping.}
\label{tab:tp}
\centering
\scriptsize
\setlength{\tabcolsep}{3pt}
\begin{tabular}{@{}l ccccc@{}}
\toprule
$N$ & 16 & 64 & 128 & 192 & 280 \\
\midrule
Non-EE (tokens/s) & 1134 & 3829 & 6335 & 7746 & 8650 \\
Within-iteration & 0.98 & 1.01 & 1.07 & 1.13 & 1.14 \\
Per-row skipping & 0.98 & 1.05 & 1.17 & 1.38 & 1.63 \\
\syshold & 1.93 & 1.96 & 2.02 & 2.09 & 2.08 \\
\sys, $\alpha = 0$ & 1.94 & 1.99 & 2.05 & 2.09 & 2.24 \\
\sys, $\alpha = .4$ & 2.12 & 2.19 & 2.16 & 2.24 & 2.35 \\
\midrule
\sys ($\alpha = .4$) / per-row & 2.16 & 2.08 & 1.85 & 1.62 & 1.44 \\
\bottomrule
\end{tabular}
\end{table}
On 13B, measured speedups exceed W's optima at $N \geq 32$ because held rows skip the deep attention, which W does not count.
The controller's choices follow the theory: on 70B at (.25, .9) its hold cap grows with $\alpha$ (from $m = 1$ at $\alpha = 0$ to $m = 3$ on 79--91\% of steps at $\alpha = .6$), and on 13B it holds ($m = 1$) on 83--100\% of steps at $N \geq 64$ and $\alpha \leq .4$, and speculates at $N = 16$ and at $\alpha = .6$.
On compute-bound 8B (\cref{tab:map8}), at full memory and $\alpha = .4$ \sys holds ($m = 1$) on 95--99\% of steps, runs $k = 1$ to 3 at (.25, .9) and $k = m = 1$ on 98--99\% of steps at (.25, .5) and (.5, .5), so it reduces to per-row skipping where that is the best policy.
Rows are not free on 8B, so W, which counts weight reads only, does not describe it.
In the cells marked $\dagger$, compact mode measures from 0.02 below to 0.05 above \syshold.
Where nothing exits, compact mode costs 5.5--9.7\% on 8B (in place: 2.8--4.2\%), so an 8B deployment whose model rarely exits runs in place.

\begin{table*}[t]
\caption{Decode throughput of the deployed systems over non-EE with the synthetic ramp at $N = 16$ / full memory (47 on 70B, 80 on 13B), sweeping the guess acceptance $\alpha$.
$\dagger$: the deployment runs in place, the calibrated model's choice at the cell's $N$, $p$, and $\alpha$, where \sys holds, so the cell repeats the number of \syshold (compact mode's in \cref{sec:appendix-grid}). $\ddagger$: the full-depth fallback ran (\cref{sec:appendix-layerskip}). Bold: $\alpha = .4$ where $p \geq E/L$. Ratios in the text use unrounded values.}
\label{tab:map}
\centering
\scriptsize
\setlength{\tabcolsep}{2.6pt}
\begin{tabular}{@{}l ccccc ccccc@{}}
\toprule
 & \multicolumn{5}{c}{\textbf{Llama-2-70B} (weight-bound)} & \multicolumn{5}{c}{\textbf{Llama-2-13B} (attention-bound)} \\
\cmidrule(lr){2-6}\cmidrule(lr){7-11}
$(E/L, p)$ & holding only & \sys, $\alpha$ 0 & $\alpha$ .2 & $\alpha$ .4 & $\alpha$ .6 & holding only & \sys, $\alpha$ 0 & $\alpha$ .2 & $\alpha$ .4 & $\alpha$ .6 \\
\midrule
(.25, .9) & 2.01 / 2.05 & 2.01 / 2.07 & 2.06 / 2.10 & \textbf{2.22 / 2.24} & 2.41 / 2.45 & 2.02 / 2.38 & 2.02$^\dagger$ / 2.38$^\dagger$ & 2.01 / 2.38$^\dagger$ & \textbf{2.13} / 2.38$^\dagger$ & 2.24 / 2.53 \\
(.375, .7) & 1.22 / 1.25 & 1.22 / 1.24 & 1.30 / 1.30 & \textbf{1.38 / 1.38} & 1.50 / 1.51 & 1.29 / 1.49 & 1.29 / 1.50 & 1.29 / 1.50 & \textbf{1.39 / 1.50} & 1.48 / 1.55 \\
(.25, .5) & 1.18 / 1.20 & 1.18 / 1.18 & 1.25 / 1.24 & \textbf{1.39 / 1.39} & 1.57 / 1.62 & 1.22 / 1.37 & 1.22 / 1.38 & 1.22 / 1.39 & \textbf{1.34 / 1.39} & 1.47 / 1.51 \\
(.5, .5) & 1.00 / 1.02 & 1.00 / 1.00 & 1.05 / 1.05 & \textbf{1.12 / 1.12} & 1.19 / 1.19 & 1.04 / 1.17 & 1.04 / 1.16 & 1.04 / 1.16 & \textbf{1.11 / 1.16} & 1.17 / 1.21 \\
(.75, .3) & 1.00 / 1.00 & 1.00$^\dagger$ / 1.00$^\dagger$ & 1.00$^\dagger$ / 1.00$^\dagger$ & 1.00$^\dagger$ / 1.00$^\dagger$ & 1.00$^\dagger$ / 1.00$^\dagger$ & 1.00 / 1.04 & 1.00$^\dagger$ / 1.03 & 1.00$^\dagger$ / 1.03 & 1.00$^\dagger$ / 1.03 & 1.00$^\dagger$ / 1.03 \\
(.5, 0) & 0.99 / 0.99 & 0.99$^\dagger$ / 0.99$^\dagger$ & 0.99$^\dagger$ / 0.99$^\dagger$ & 0.99$^\dagger$ / 0.99$^\dagger$ & 1.02 / 1.02 & 0.98$^\ddagger$ / 0.99$^\ddagger$ & 0.98$^{\dagger\ddagger}$ / 0.99$^{\dagger\ddagger}$ & 0.98$^{\dagger\ddagger}$ / 0.99$^{\dagger\ddagger}$ & 0.98$^{\dagger\ddagger}$ / 0.99$^{\dagger\ddagger}$ & 0.98 / 0.99$^{\dagger\ddagger}$ \\
\bottomrule
\end{tabular}
\end{table*}

\begin{table*}[t]
\caption{Llama-3.1-8B (compute-bound) with the synthetic ramp: decode throughput of the deployed systems over non-EE at $N = 16$ / full memory (527) and at the middle concurrency $N = 384$ ($\dagger$, $\ddagger$, and bold as in \cref{tab:map}).}
\label{tab:map8}
\centering
\scriptsize
\setlength{\tabcolsep}{2.6pt}
\begin{tabular}{@{}l ccccc ccccc@{}}
\toprule
 & \multicolumn{5}{c}{$N = 16$ / full memory} & \multicolumn{5}{c}{$N = 384$} \\
\cmidrule(lr){2-6}\cmidrule(lr){7-11}
$(E/L, p)$ & holding only & \sys, $\alpha$ 0 & $\alpha$ .2 & $\alpha$ .4 & $\alpha$ .6 & hold & $\alpha$ 0 & .2 & .4 & .6 \\
\midrule
(.25, .9) & 1.37 / 1.86 & 1.37$^\dagger$ / 1.95 & 1.37$^\dagger$ / 1.97 & \textbf{1.33 / 1.97} & 1.37 / 1.97 & 1.79 & 1.85 & 1.86 & 1.87 & 1.89 \\
(.375, .7) & 1.13 / 1.39 & 1.13$^\dagger$ / 1.42 & 1.13 / 1.43 & \textbf{1.16 / 1.43} & 1.23 / 1.43 & 1.37 & 1.41 & 1.40 & 1.40 & 1.41 \\
(.25, .5) & 0.97$^\ddagger$ / 1.21 & 0.97$^{\dagger\ddagger}$ / 1.32 & 0.97$^{\dagger\ddagger}$ / 1.32 & \textbf{1.13 / 1.32} & 1.15 / 1.32 & 1.19 & 1.27 & 1.27 & 1.27 & 1.27 \\
(.5, .5) & 0.98$^\ddagger$ / 1.12 & 0.98$^{\dagger\ddagger}$ / 1.17 & 0.98$^{\dagger\ddagger}$ / 1.18 & \textbf{1.03 / 1.18} & 1.09 / 1.17 & 1.11 & 1.16 & 1.15 & 1.15 & 1.15 \\
(.75, .3) & 0.98$^\ddagger$ / 1.01 & 0.98$^{\dagger\ddagger}$ / 1.00 & 0.98$^{\dagger\ddagger}$ / 1.00 & 0.98$^{\dagger\ddagger}$ / 1.00 & 0.98$^{\dagger\ddagger}$ / 1.00 & 1.01 & 1.01 & 1.00 & 1.00 & 1.00 \\
(.5, 0) & 0.97$^\ddagger$ / 0.96 & 0.97$^{\dagger\ddagger}$ / 0.96$^\dagger$ & 0.97$^{\dagger\ddagger}$ / 0.96$^\dagger$ & 0.97$^{\dagger\ddagger}$ / 0.96$^\dagger$ & 0.97$^{\dagger\ddagger}$ / 0.96$^\dagger$ & 0.96 & 0.96$^\dagger$ & 0.96$^\dagger$ & 0.96$^\dagger$ & 0.96$^\dagger$ \\
\bottomrule
\end{tabular}
\end{table*}

\begin{table*}[t]
\caption{Within-iteration batching and per-row skipping with the synthetic ramp: decode throughput over non-EE at $N = 16$ / full memory (47 on 70B, 80 on 13B, 527 on 8B), to three decimals because the gains are small.}
\label{tab:map-base}
\centering
\scriptsize
\setlength{\tabcolsep}{4pt}
\begin{tabular}{@{}l cc cc cc@{}}
\toprule
 & \multicolumn{2}{c}{\textbf{Llama-2-70B} (weight-bound)} & \multicolumn{2}{c}{\textbf{Llama-2-13B} (attention-bound)} & \multicolumn{2}{c}{\textbf{Llama-3.1-8B} (compute-bound)} \\
\cmidrule(lr){2-3}\cmidrule(lr){4-5}\cmidrule(lr){6-7}
$(E/L, p)$ & within-iteration & per-row skipping & within-iteration & per-row skipping & within-iteration & per-row skipping \\
\midrule
(.25, .9) & 1.000 / 1.038 & 1.011 / 1.070 & 1.187 / 1.683 & 1.177 / 1.720 & 0.974 / 1.521 & 0.932 / 1.939 \\
(.375, .7) & 1.006 / 1.031 & 1.004 / 1.042 & 1.108 / 1.355 & 1.095 / 1.348 & 0.950 / 1.270 & 0.913 / 1.438 \\
(.25, .5) & 0.992 / 1.017 & 0.993 / 1.032 & 1.074 / 1.281 & 1.070 / 1.283 & 0.954 / 1.216 & 0.913 / 1.330 \\
(.5, .5) & 1.004 / 1.019 & 0.990 / 1.015 & 1.045 / 1.167 & 1.040 / 1.167 & 0.953 / 1.125 & 0.917 / 1.184 \\
(.75, .3) & 0.999 / 1.003 & 0.995 / 1.000 & 1.001 / 1.038 & 0.997 / 1.036 & 0.948 / 1.014 & 0.921 / 1.012 \\
(.5, 0) & 0.997 / 0.997 & 0.996 / 0.997 & 0.974 / 0.988 & 0.975 / 0.989 & 0.935 / 0.965 & 0.902 / 0.939 \\
\bottomrule
\end{tabular}
\end{table*}

\section{Tail Latency in Detail}
\label{sec:appendix-tails}

\begin{figure}[t]
\centering
\begin{tikzpicture}[font=\footnotesize]
\begin{groupplot}[
  group style={group size=2 by 2, horizontal sep=0.9cm, vertical sep=1.55cm},
  width=0.53\columnwidth,
  height=2.8cm,
  xlabel={p99 max gap (ms)},
  flexit axis,
]
\nextgroupplot[title={70B, (.25, .9), $N = 32$}, ylabel={Speedup}, xmin=30, xmax=125, ymin=1.4, ymax=2.55,
  legend to name=gaplegend, legend columns=3, flexit legend,
  legend style={/tikz/every even column/.append style={column sep=0.6ex}}]
  \draw[black!45, densely dotted] (axis cs:53.5,1.4) -- (axis cs:53.5,2.55);
  \draw[black!45, densely dotted] (axis cs:40.2,1.4) -- (axis cs:40.2,2.55);
  \addplot[sys hold] coordinates {(69.8,2.022) (47.5,1.834) (35.0,1.525)};
  \addlegendentry{holding only}
  \addplot[sys flexit] coordinates {(91.8,2.258) (49.4,1.856) (40.2,1.524)};
  \addlegendentry{\sys, $\alpha{=}.4$}
  \addplot[sys flexit, color=sysflexit!55, mark=o] coordinates {(116.4,2.414) (49.3,1.885) (42.7,1.544)};
  \addlegendentry{\sys, $\alpha{=}.6$}
\nextgroupplot[title={70B, (.25, .9), $N = 47$ (full)}, xmin=30, xmax=125, ymin=1.4, ymax=2.55]
  \draw[black!45, densely dotted] (axis cs:56.1,1.4) -- (axis cs:56.1,2.55);
  \draw[black!45, densely dotted] (axis cs:42.1,1.4) -- (axis cs:42.1,2.55);
  \addplot[sys hold] coordinates {(68.1,2.047) (49.3,1.847) (36.6,1.535)};
  \addplot[sys flexit] coordinates {(93.7,2.232) (50.8,1.823) (39.8,1.484)};
  \addplot[sys flexit, color=sysflexit!55, mark=o] coordinates {(102.2,2.433) (51.7,1.818) (39.4,1.490)};
\nextgroupplot[title={13B, (.25, .9), $N = 64$}, ylabel={Speedup}, xmin=15, xmax=40, ymin=2.05, ymax=2.4]
  \draw[black!45, densely dotted] (axis cs:28.1,2.05) -- (axis cs:28.1,2.4);
  \draw[black!45, densely dotted] (axis cs:21.1,2.05) -- (axis cs:21.1,2.4);
  \addplot[sys hold] coordinates {(33.7,2.337) (24.9,2.305) (18.7,2.168)};
  \addplot[sys flexit] coordinates {(38.0,2.328) (25.0,2.264) (19.7,2.134)};
\nextgroupplot[title={8B, (.25, .5), $N = 16$}, xmin=4.5, xmax=9.2, ymin=0.9, ymax=1.2]
  \draw[black!45, densely dotted] (axis cs:7.8,0.9) -- (axis cs:7.8,1.2);
  \draw[black!45, densely dotted] (axis cs:5.8,0.9) -- (axis cs:5.8,1.2);
  \addplot[sys hold] coordinates {(5.71,0.978) (5.30,0.964) (5.68,0.964)};
  \addplot[sys flexit] coordinates {(8.80,1.092) (7.55,0.951) (7.50,0.945)};
\end{groupplot}
\node[anchor=south] at ($(group c1r1.north)!0.5!(group c2r1.north)+(0,0.3cm)$) {\pgfplotslegendfromname{gaplegend}};
\end{tikzpicture}
\caption{Gap bounds, speedup over non-EE: each line runs from unbounded to 2$\times$ and 1.5$\times$ non-EE's TPOT (dotted, 2$\times$ right, 1.5$\times$ left).
 8B at $N \geq 384$ is not drawn: one 0.26--0.28\,s pause per run, non-EE's included, exceeds any such bound there.}
\label{fig:gapbound}
\end{figure}

\heading{Holding trades tail gaps for TPOT, and a gap bound caps them}
Where EE pays, \syshold and \sys cut the TPOT to 0.40--0.85$\times$ non-EE's, but held and tentative tokens wait for their flush, so without a bound the median p99 largest gap is 1.29$\times$ non-EE's on 13B and 1.72$\times$ on 70B for \syshold, and 1.61$\times$ and 2.64$\times$ for \sys at $\alpha = .4$.
At the 70B headline cell, bounds of 2$\times$ and 1.5$\times$ non-EE's TPOT hold at p99 in 11 of 12 cells and within one step in the other (\cref{fig:gapbound}), and \sys keeps 0.75--0.82 and 0.61--0.68 of its unbounded throughput, holding 0.90--0.91 and 0.75.
The tight bound cuts windows to about four steps, where speculation no longer pays for its wider deep passes, so \sys runs at 0.97--1.03$\times$ \syshold, against 1.09--1.19$\times$ unbounded.
On 13B at (.25, .9) with $N = 64$, both bounds hold at p99 in all four cells, \syshold keeps 0.986 and 0.928 of its throughput and \sys 0.973 and 0.917, so under a bound \sys runs 1--2\% below \syshold, as on 70B.

\Cref{tab:tails} gives the latency of every arm over the cells where EE pays.
Compact arms add two sources of delay to \syshold: a request that an undersized bucket leaves to the next flush, and \sys's probe and rejected windows.
Free-row sizing raises each flush's buckets to the largest size below the calibrated ridge.
In a same-node ablation of \syscphold over the cells where EE pays, free rows keep its throughput (median 1.000) and cut its p99 largest gap to a median 0.72$\times$. Per-row skipping runs with free rows off, its best-throughput setting.

\heading{The gap clock}
The gap rule forces a flush when waiting one more step would let the oldest held request's gap pass the bound, and with a one-step margin it does so one shallow period earlier.
On 13B at (.25, .9) with $N = 64$ and bounds of 3$\times$, 2$\times$, and 1.5$\times$ non-EE's TPOT, the margin removes every gap above the bound but those of one 23\,ms host stall (0.006\% of gaps at 3$\times$, against 0.07--0.30\% without it) at a cost of up to 3.9 points of unbounded throughput.

\begin{table*}[t]
\caption{Latency over non-EE where EE pays (median p99 largest gap in parentheses). On 13B, four cells include one 78--95\,ms GPU step that every request in flight waits on. On 8B, gaps at $N \geq 384$ include one batch-wide pause of 0.26--0.28\,s per run that non-EE has too.}
\label{tab:tails}
\centering
\scriptsize
\setlength{\tabcolsep}{3pt}
\begin{tabular}{@{}l cc cc cc@{}}
\toprule
 & \multicolumn{2}{c}{\textbf{13B}} & \multicolumn{2}{c}{\textbf{70B}} & \multicolumn{2}{c}{\textbf{8B}} \\
\cmidrule(lr){2-3}\cmidrule(lr){4-5}\cmidrule(lr){6-7}
Arm & TPOT & p99 gap & TPOT & p99 gap & TPOT & p99 gap \\
\midrule
Within-iter. & 0.59--0.93 & 0.56--0.96 (0.80) & 0.96--1.01 & 0.93--1.00 (0.98) & 0.66--1.05 & 0.98--1.07 (1.04) \\
Holding only & 0.42--0.82 & 1.08--3.80 (1.29) & 0.49--0.85 & 1.23--2.45 (1.72) & 0.54--1.03 & 1.00--2.33 (1.06) \\
Per-row skip. & 0.58--0.93 & 0.72--1.47 (0.92) & 0.93--1.01 & 0.95--1.95 (1.83) & 0.52--1.10 & 0.98--1.09 (1.03) \\
Comp., holding only & 0.41--0.82 & 1.07--1.98 (1.29) & 0.48--0.85 & 1.35--2.95 (1.72) & 0.51--1.09 & 1.01--2.63 (1.02) \\
\sys, $\alpha = 0$ & 0.41--0.82 & 1.12--4.60 (1.69) & 0.48--0.85 & 1.38--2.72 (1.75) & 0.51--1.09 & 1.00--3.23 (1.04) \\
\sys, $\alpha = .4$ & 0.42--0.75 & 1.06--4.88 (1.61) & 0.44--0.73 & 1.76--4.13 (2.64) & 0.51--0.88 & 1.01--3.39 (1.05) \\
\sys, $\alpha = .6$ & 0.40--0.68 & 1.64--11.15 (2.19) & 0.41--0.67 & 2.03--4.08 (3.13) & 0.51--0.87 & 1.03--4.65 (1.06) \\
\bottomrule
\end{tabular}
\end{table*}

\section{Open-Loop Serving}
\label{sec:appendix-openloop}

\begin{figure}[t]
\centering
\begin{tikzpicture}[font=\footnotesize]
\begin{groupplot}[
  group style={group size=2 by 1, horizontal sep=0.3cm, y descriptions at=edge left},
  width=0.53\columnwidth,
  height=3.4cm,
  xmin=0.7, xmax=4.3, xtick={1,2,3,4},
  ymin=0, ymax=4.2, ytick={0,1,2,3,4},
  xlabel={Offered rate (req/s)},
  flexit axis,
]
\nextgroupplot[title={(.25, .9)}, ylabel={Goodput (req/s)},
  legend to name=openlegend, legend columns=2, flexit legend]
  \addplot[sys nonee] coordinates {(1,0.78) (2,1.59) (3,0.95) (4,0.26)};
  \addlegendentry{non-EE}
  \addplot[sys hold] coordinates {(1,0.83) (2,1.70) (3,2.60) (4,3.55)};
  \addlegendentry{\sys (holding only)}
  \addplot[sys flexit] coordinates {(1,0.83) (2,1.70) (3,2.62) (4,3.58)};
  \addlegendentry{\sys}
  \addplot[sys flexit, color=sysflexit!55, dashed, mark=o, mark options={solid}] coordinates {(1,0.83) (2,1.59) (3,0.69) (4,0.27)};
  \addlegendentry{\sys, 70\,ms bound}
\nextgroupplot[title={(.25, .5)}]
  \addplot[sys nonee] coordinates {(1,0.78) (2,1.59) (3,0.95) (4,0.26)};
  \addplot[sys hold] coordinates {(1,0.79) (2,1.62) (3,2.25) (4,0.35)};
  \addplot[sys flexit] coordinates {(1,0.80) (2,1.64) (3,2.50) (4,0.36)};
  \addplot[sys flexit, color=sysflexit!55, dashed, mark=o, mark options={solid}] coordinates {(1,0.80) (2,1.58) (3,0.43) (4,0.21)};
\end{groupplot}
\node[anchor=south] at ($(group c1r1.north)!0.5!(group c2r1.north)+(0,0.3cm)$) {\pgfplotslegendfromname{openlegend}};
\end{tikzpicture}
\caption{Open loop on Llama-2-70B ($\alpha = .4$): requests/s that meet TTFT $\leq$ 2\,s and TPOT $\leq$ 50\,ms.}
\label{fig:openloop}
\end{figure}

We serve 70B at $\alpha = .4$ at Poisson rates of 1--4 requests/s over the five tasks' prompts and 1000 output tokens, allow preemption, and count goodput under (TTFT, TPOT) bounds of (1\,s, 40\,ms), (2\,s, 50\,ms), and (5\,s, 75\,ms).

\heading{The decode gain becomes serving capacity}
At (.25, .9), non-EE saturates between 2 and 3 requests/s, where 27--61\% of its requests meet the bounds, and 7--14\% at 4 requests/s (\cref{fig:openloop}).
\syshold and \sys serve 4 requests/s with every request within all three bounds (TTFT p99 $\leq$ 0.40\,s).
At (.25, .5) and 3 requests/s, \sys serves 99--100\% of requests within the bounds, \syshold 92--99\%, and non-EE 27--61\%, and at 4 requests/s every system is overloaded (7--22\%).
At 1--2 requests/s and (.25, .9), the EE systems cut the TPOT p50 from non-EE's 28--31\,ms to 12--15\,ms at a TTFT within 15\,ms of non-EE's, and where a system meets the bounds, tokens generated after a request's first resume are 0--1.5\% of the output for \sys and 0--7.6\% for holding, against 26--40\% for every overloaded system, non-EE included.
Prefill stalls (non-EE's p99 gap: 215--342\,ms) overload the 70\,ms gap bound from 3 requests/s, so a bound must sit above them.

\heading{13B and 8B}
We scale the request rates by each model's non-EE decode capacity at full memory (2.9$\times$ and 16.9$\times$ 70B's: 3--12 and 17--68 requests/s), and the TPOT bounds and the gap bound by its non-EE TPOT, over the same prompts (13B: 60\,s of arrivals per cell, 8B: all 2,312 prompts).
On 13B at (.25, .9), non-EE serves 49\% of requests within the bounds at 9 requests/s and 27\% at 12, while \syshold and \sys serve 12 requests/s with every request within all three bounds (the gap-bounded \sys 80--86\%), and at (.25, .5) \syshold serves 9 requests/s in full and \sys 80--100\%.
On 8B at (.25, .9) and 34 requests/s, non-EE serves 57--73\% of requests within the bounds, while \syshold and compact \sys serve every request, at 1.8--2.3$\times$ non-EE's goodput (29--30 against 13--17 requests/s).
At (.25, .5), \syshold serves 88--92\% and compact \sys 70--71\%, which preempts more requests than in-place holding (114 against 79), the open-loop face of the compaction choice that errs on 8B below full memory (\cref{sec:appendix-sd}), and from 51 requests/s every system is overloaded.

\section{Overheads and Ablations}
\label{sec:appendix-overhead}

\begin{table}[t]
\caption{Ablations: throughput with the element over without it.}
\label{tab:ablation}
\centering
\scriptsize
\setlength{\tabcolsep}{3pt}
\begin{tabular}{@{}l r r r@{}}
\toprule
\textbf{Element} & \textbf{13B} & \textbf{70B} & \textbf{8B} \\
\midrule
Speculation, $\alpha = .4$ & & 1.08--1.17 & 0.97--1.17 \\
Speculation, $\alpha = .6$ & & 1.18--1.36 & 1.00--1.19 \\
Speculation, $\alpha = 0$ & & 0.989--1.003 & 0.975--1.000 \\
No host sync (plan-ahead) & 1.37--1.43 & 1.08--1.11 & 1.34--1.69 \\
\quad non-EE's own overlap & 1.11--1.14 & 1.05 & 1.28--1.36 \\
Controller vs.\ best fixed $(k, m)$ & 0.975--1.007 & 0.993--1.002 & 0.920--1.077 \\
Compaction (holding only) & \multicolumn{2}{r}{0.984--1.019} & 1.033--1.097 \\
\bottomrule
\end{tabular}
\end{table}

\heading{Each element earns its place}
Synchronizing the host every step costs non-EE 10.2--12.3\%, 4.3\%, and 22--26\% on 13B, 70B, and 8B, less than it costs the EE engines (\cref{tab:ablation}), so plan-ahead flushing recovers more than non-EE's own overlap.
The controller, never told $\alpha$, is within 0.7\% of the best fixed $(k, m)$ on 70B and 2.5\% on 13B, where it holds while $(3, 2)$ is 2.5\% faster.
On 8B it is within 8.0\%: at (.25, .9) and $N = 16$ a fixed $(3, 2)$ runs 8.7\% faster than its $k = 4$ and 5, while at full memory the controller runs 1.1\% faster than every fixed $(k, m)$ at (.25, .9) and 7.7\% faster at (.25, .5), where it runs $k = 1$, which no fixed candidate does.
Compaction pays only on 8B.

Same-node repeats of one engine agree within 0.8\% (median 0.15\%), each engine's mean is within 0.5\% of its run in the grid on another node, and arm ratios agree within 0.9\%.

\section{Grouped Policies at Their Best}
\label{sec:appendix-grouped}

Majority and Greedy force 17--30\% of tokens out (\cref{tab:involuntary}) and lose 30 and 34 points on BBH.
Our grouped engines read the reduced decision on the host every step, so we also re-cost their logged decisions without that synchronization, with exit steps at \sys's shallow-step time and stay steps at non-EE's (\cref{tab:grouped}).
At zero overhead, Consensus is 1.00$\times$ non-EE at full memory in 10 of 12 cells (13B with the exit after layer 15 and $\epsilon^*$ 1\% / 3\%: 1.03 / 1.17), and 1.44--2.09$\times$ on 13B at $N = 16$, where all 16 tokens often exit together, against a measured 0.94--1.02$\times$ on 70B and 0.47--0.92$\times$ on 13B.
It forces about half of the tokens to stay (\cref{tab:involuntary}), so it serves nearly the non-EE model.
Greedy and Majority would run faster than \sys on greedy PG-19, where LayerSkip wants to exit almost every token, but only by forcing the rest out (\cref{tab:involuntary}).

\begin{table}[t]
\caption{Grouped policies on greedy PG-19 at $\epsilon^*$ 3\%, exit after layer 10, $N$ = 16 / mid / full memory: measured and re-costed zero-overhead speedup over non-EE. Greedy continuations of PG-19 fall into repetition loops (distinct-4 of .04--.05) and LayerSkip's exit rate climbs toward .98, so these cells are a high-$p$ stress test that we do not count for early exit.}
\label{tab:grouped}
\centering
\footnotesize
\setlength{\tabcolsep}{2.4pt}
\begin{tabular}{@{}l l c c@{}}
\toprule
Model & Policy & Measured & Zero-overhead \\
\midrule
13B & Greedy & 1.11 / 1.57 / 1.71 & 3.53 / 3.83 / 3.51 \\
13B & Majority & 1.07 / 1.53 / 1.68 & 3.40 / 3.78 / 3.46 \\
70B & Greedy & 2.69 / 2.70 / 2.75 & 7.59 / 7.67 / 7.54 \\
70B & Majority & 2.53 / 2.61 / 2.60 & 6.14 / 6.32 / 6.02 \\
\bottomrule
\end{tabular}
\end{table}

\section{LayerSkip Methodology}
\label{sec:appendix-layerskip}

\heading{Depths and thresholds}
A teacher-forced sweep over the non-EE model's outputs records, at exit depths $E/L \in \{.125, .25, .375, .5\}$, the ramp's margin and whether the exit layer's argmax agrees with the non-EE model's, and at $E/L \geq .25$ the margin separates agreement with an AUROC of 0.79--0.98 on 13B and 0.87--0.97 on 70B.
Thresholds come from a calibration split, one per exit layer, task, and target (an exit rate $p^*$ or a disagreement rate $\epsilon^*$ among exited tokens), and every request applies its task's threshold.
One live recalibration then reset each exit-rate threshold whose live run missed $p^*$ by more than .05 to the threshold that hits $p^*$ on that run's logged margins over the calibration documents, risk targets keep their teacher-forced thresholds, and we report realized exit rates on held-out documents.
Quality runs use natural lengths at $N = 64$ with exits after layers 10 and 15 on 13B, 10 and 20 on 70B, and 8 and 12 on 8B, against the non-EE model (A) and batch-1 EE (B) in a reference implementation, with 95\% paired-bootstrap intervals over documents.
The held-out sets are 250 GSM8K, 500 BBH, 249 MBPP, 82 HumanEval, and 150 CNN/DM documents on 13B, 100 GSM8K documents on 70B, and 250 GSM8K documents on 8B, scored by exact match, pass@1, and ROUGE-2.

\begin{table}[t]
\caption{Quality at $N = 64$, the ranges behind \cref{fig:quality}: non-EE model (A), ranges over two depths and three targets of the EE model's loss (B $-$ A, B = batch-1 EE) and of \sys $-$ B, and its 95\% intervals that contain 0 ($\dagger$: the other favours \sys).}
\label{tab:quality}
\centering
\footnotesize
\setlength{\tabcolsep}{2.6pt}
\begin{tabular}{@{}l c c c c@{}}
\toprule
\textbf{Task} (model) & \textbf{A} & \textbf{B $-$ A} & \textbf{\sys $-$ B} & \textbf{0 in CI} \\
\midrule
GSM8K (13B) & .264 & $-$.18 to $-$.05 & $-$.020 to +.012 & 6/6 \\
BBH (13B) & .468 & $-$.26 to $-$.04 & $-$.006 to +.018 & 5/6$^\dagger$ \\
MBPP (13B) & .281 & $-$.18 to $-$.08 & $-$.004 to +.008 & 6/6 \\
HumanEval (13B) & .098 & $-$.06 to +.01 & .000 & 6/6 \\
CNN/DM (13B, pts) & 16.4 & $-$0.4 to +0.8 & $-$0.4 to +0.6 & 5/6$^\dagger$ \\
GSM8K (70B) & .490 & $-$.29 to $-$.14 & .00 to +.04 & 6/6 \\
GSM8K (8B) & .440 & $-$.33 to $-$.19 & $-$.020 to +.016 & 6/6 \\
\bottomrule
\end{tabular}
\end{table}

\heading{Summarization runs}
\Cref{fig:map}b and \cref{tab:summ} come from one job per model on held-out CNN/DM documents in LayerSkip's 1-shot format, with natural lengths up to 128 tokens and greedy decoding.
Full-memory runs decode 1,600 documents and $N = 16$ and 64 the first 400.
Thresholds for CNN/DM come from 400 calibration documents, and the realized held-out exit rates lie 2--5 points below the calibrated ones.
Each batch of $N$ requests is released at once and run to completion, and the scheduler holds decode steps until its last prompt is prefilled, so the batch decodes together, as in the synthetic runs' static batch, and \sys, holding, and EAGLE are each measured over their own model's non-EE run in the same job.
Throughput runs from the last prompt's first token to the batch's last token, so they include the drain of a batch at natural lengths.

\begin{table*}[t]
\caption{CNN/DM summarization at $N$ = 16 / 64 / full memory: decode speedup over non-EE over each batch's whole decode phase, and the realized exit rate at full memory ($\epsilon^*$ 1\% / 3\%). 13B exits at layer 15, 70B at 20, and 8B at 8. EAGLE at its best $k \leq 4$ over its chat target, on 8B with the better of two EAGLE-3 heads. Ratios in the text use unrounded values.}
\label{tab:summ}
\centering
\scriptsize
\setlength{\tabcolsep}{4.5pt}
\begin{tabular}{@{}l ccc@{}}
\toprule
System & \multicolumn{3}{c}{\textbf{CNN/DM}} \\
\cmidrule(lr){2-4}
 & 13B & 70B & 8B \\
\midrule
\sys, $\tau = \infty$ & 1.27 / 1.30 / 1.27 & 1.42 / 1.43 / 1.43 & 0.95 / 0.91 / 0.92 \\
\sys, $\epsilon^*$ 1\% & 1.70 / 1.91 / 1.86 & 1.58 / 1.60 / 1.58 & 0.93 / 0.86 / 1.91 \\
\sys, $\epsilon^*$ 3\% & 1.53 / 1.78 / 1.78 & 1.57 / 1.64 / 1.59 & 1.06 / 1.11 / 1.96 \\
Holding only, $\epsilon^*$ 1\% & 1.63 / 1.82 / 1.85 & 1.25 / 1.29 / 1.26 & 0.91 / 0.92 / 1.71 \\
Holding only, $\epsilon^*$ 3\% & 1.54 / 1.79 / 1.79 & 1.50 / 1.50 / 1.50 & 1.06 / 1.24 / 1.88 \\
Per-row skip., $\epsilon^*$ 1\% & 1.31 / 1.53 / 1.67 & 1.08 / 1.12 / 1.12 & 0.95 / 1.12 / 1.92 \\
Per-row skip., $\epsilon^*$ 3\% & 1.16 / 1.43 / 1.58 & 1.07 / 1.14 / 1.15 & 0.97 / 1.14 / 2.14 \\
Within-iter., $\epsilon^*$ 1\% & 1.31 / 1.52 / 1.66 & 1.09 / 1.13 / 1.12 & 0.99 / 1.13 / 1.56 \\
Within-iter., $\epsilon^*$ 3\% & 1.16 / 1.41 / 1.56 & 1.07 / 1.13 / 1.13 & 1.01 / 1.16 / 1.64 \\
EAGLE & 1.73 / 1.61 / 1.58 & 1.68 / 1.43 / 1.45 & 1.48 / 1.32 / 1.17 \\
\midrule
Exit rate & .88 / .93 & .58 / .84 & .87 / .93 \\
\bottomrule
\end{tabular}
\end{table*}

\Cref{tab:summ-abs} gives the absolute numbers behind \cref{fig:map}b: \sys decodes 1.23--1.61$\times$ EAGLE's tokens/s, and \sys's absolute ROUGE-2 lies $-$0.8 to +1.1 points from EAGLE's.
Compact \sys on 8B fits 795 requests against 813 for the other systems.
On 8B below full memory, no EE system gains much over non-EE, because its steps are short, so per-step costs that non-EE does not pay, the ramp's LM head (6.5\% of 8B's weights) and the host and sampling time of \cref{sec:appendix-costmodel}, take a larger share of each step, and at $N = 64$ the compact pick runs 7\% below \syshold (\cref{sec:appendix-sd}).

\begin{table}[t]
\caption{Absolute values behind \cref{fig:map}b (CNN/DM, full memory, $\epsilon^*$ 1\%, 13B at layer 15): decode throughput (thousand tokens/s) and ROUGE-2 (points).}
\label{tab:summ-abs}
\centering
\scriptsize
\setlength{\tabcolsep}{2.4pt}
\begin{tabular}{@{}l ccc@{}}
\toprule
System & 13B ($N$ 104--105) & 70B ($N$ 61) & 8B ($N$ 795--813) \\
\midrule
Non-EE, LayerSkip & 3.79 / 16.5 & 1.75 / 19.1 & 26.6 / 17.6 \\
\sys, $\tau = \infty$ & 4.81 / 16.5 & 2.50 / 19.1 & 24.3 / 17.7 \\
\sys, $\epsilon^*$ 1\% & 7.07 / 17.5 & 2.75 / 18.5 & 50.8 / 16.2 \\
Holding only, $\epsilon^*$ 1\% & 6.99 / 17.5 & 2.21 / 18.6 & 45.4 / 16.2 \\
\midrule
Non-EE, chat target & 3.56 / 16.6 & 1.54 / 19.3 & 26.9 / 16.4 \\
EAGLE, best $k$ & 5.64 / 16.4 & 2.24 / 19.3 & 31.5 / 16.4 \\
\bottomrule
\end{tabular}
\end{table}

\heading{Sampled PG-19 and coupled guessing}
\Cref{tab:layerskip-speed} lists every sampled cell as the mean of three repeats on one node.
On a tiny model (temperature 1 with top-p 1 and .95, and temperature .7) and a 1B LayerSkip model (temperature 1, top-p .95), speculating engines with hold caps 2 and 4 produced the same tokens as a non-speculating reference engine for all 17 requests of each test.

\begin{table}[t]
\caption{LayerSkip on sampled PG-19 at $N$ = 16 / mid / full memory (mid: 64 on 13B, 32 on 70B, 256 on 8B, full: 86, 47, 528), means of three repeats: exit rate $p$ of \syshold, \syshold and \sys over non-EE, and \sys's live acceptance ($\ddagger$: the full-depth fallback ran, \cref{sec:appendix-layerskip}).}
\label{tab:layerskip-speed}
\centering
\scriptsize
\setlength{\tabcolsep}{1.6pt}
\begin{tabular}{@{}l l c c c c@{}}
\toprule
Model & Target & $p$ & Holding only & \sys & $\alpha$ \\
\midrule
13B & $\epsilon^*$ 1\% & .28--.29 & 1.02 / 1.11 / 1.13 & 1.16 / 1.11 / 1.17 & .39 / .39 / .42 \\
13B & $\epsilon^*$ 3\% & .49--.51 & 1.20 / 1.40 / 1.51 & 1.32 / 1.42 / 1.49 & .41 / .40 / .37 \\
13B & $p^*$ .5 & .05--.06 & 0.97$^\ddagger$ / 1.00 / 1.01 & 1.01 / 0.97 / 0.99 & .41 / .39 / .40 \\
70B & $\epsilon^*$ 1\% & .31--.34 & 1.02 / 1.05 / 1.05 & 1.20 / 1.16 / 1.18 & .28 / .31 / .31 \\
70B & $\epsilon^*$ 3\% & .51--.56 & 1.21 / 1.17 / 1.18 & 1.32 / 1.28 / 1.32 & .29 / .28 / .27 \\
70B & $p^*$ .5 & .06--.07 & 0.99 / 0.99 / 0.99 & 1.08 / 1.07 / 1.07 & .38 / .38 / .38 \\
8B & $\epsilon^*$ 1\% & .17--.29 & 0.96$^\ddagger$ / 1.04 / 1.08 & 0.96 / 1.02 / 1.07 & .43 / .41 / .41 \\
8B & $\epsilon^*$ 3\% & .55--.57 & 1.06$^\ddagger$ / 1.10 / 1.16 & 1.00 / 1.12 / 1.28 & .38 / .34 / .36 \\
8B & $p^*$ .5 & .01--.02 & 0.96$^\ddagger$ / 0.94 / 0.97 & 0.91 / 0.93 / 0.94 & .51 / .52 / .51 \\
\bottomrule
\end{tabular}
\end{table}

\heading{Same-engine noise}
Three repeats of each LayerSkip engine on one node spread by a median of 1.5--1.9\% on 13B and 2.5--2.6\% on 70B (0.15\% on 8B), against at most 0.6\% for synthetic engines, so a single run resolves the 70B gain but not small 13B differences.
The spread follows the realized exit rate, which moves between repeats because margins near the threshold flip with batch-dependent rounding and the sampled continuations then diverge.

\heading{Live engines and synthetic twins}
Each cell's twin runs the same engine, weights, prompts, sampling, and lengths with the synthetic ramp at the exit rate measured in the cell's earlier single live run, and \sys's twin also at that run's acceptance.
Both sides are measured engine runs and no cost model is involved, so live over twin isolates the effect of LayerSkip's own exit and guess distribution against i.i.d.\ draws at the same aggregate rates, for this one model.
Live over twin lies within $\pm$5\% in 31 of 36 (cell, system) pairs, with medians of 1.00 for \sys on both models and 0.99--1.00 for \syshold, and ranges 0.957--1.115 on 13B and 0.943--1.041 on 70B.
\sys's twin at live $\alpha$ thus matches as closely as that of \syshold, so on this model the synthetic ramp's independent guesses do not overstate speculation's gain.
On LayerSkip-8B all 18 pairs lie within $\pm$5\% (0.976--1.032, median 1.00).
All five exceptions on 13B and 70B are at $\epsilon^*$ 3\%, the noisiest target: on 13B at $N \geq 64$ both systems run 5--12\% faster live than their twins, and on 70B at $N = 32$ \sys runs 5.7\% slower.

\heading{Self-speculation at $\tau = \infty$}
The self-speculation engine is \sys's compact engine with no ramp target: no position exits, every position is guessed from the exit layer's top-1 token, or its coupled draw when sampling, and verified by the deep pass with the shallow KV reused, so it emits the non-EE model's tokens, up to floating-point near-ties.
One same-node job runs non-EE, \sys at the three targets, and self-speculation under the controller and fixed windows $k = m = 2$, 3, and 4 (2 and 3 at the second exit layer) on the speed cells of \cref{tab:layerskip-speed} and on greedy PG-19, and all 136 cells are valid.
\Cref{tab:selfspec} gives the sampled cells, where the self-speculation baseline is the better of the controller and the best fixed window.

\begin{table*}[t]
\caption{Self-speculation at $\tau = \infty$ in \sys's engine on sampled PG-19 ($N$ = 16, 64 or 32, full memory): speed over non-EE of the controller and fixed windows $k = m = 2, 3, 4$, the engine's coupled acceptance (best configuration), the measured $1 - \mathrm{TV}$, the estimate from W with the coupled acceptance (which reproduces the measured best) and with $1 - \mathrm{TV}$ (rejection sampling, a model-based estimate), and \sys over measured self-speculation and over the rejection-sampling estimate at $\epsilon^*$ 1\% / 3\% / $p^*$ .5.}
\label{tab:selfspec}
\centering
\scriptsize
\setlength{\tabcolsep}{3.5pt}
\begin{tabular}{@{}l l c c c c c c c@{}}
\toprule
 & & \multicolumn{2}{c}{\textbf{Self-speculation, measured}} & & \multicolumn{2}{c}{\textbf{Estimate from W}} & \multicolumn{2}{c}{\textbf{\sys over self-speculation}} \\
\cmidrule(lr){3-4}\cmidrule(lr){6-7}\cmidrule(lr){8-9}
Model & $N$ & ctl / $k$ 2 / 3 / 4 & $\alpha$ & $1 - \mathrm{TV}$ & coupled & rejection & measured & rejection estimate \\
\midrule
13B & 16 & 0.97 / 1.03 / 0.92 / 0.80 & 0.43 & 0.50 & 1.05 & 1.09 & 1.10 / 1.35 / 0.98 & 1.05 / 1.28 / 0.93 \\
13B & 64 & 0.97 / 1.08 / 0.84 / 0.73 & 0.43 & 0.48 & 1.07 & 1.11 & 1.05 / 1.31 / 0.90 & 1.02 / 1.27 / 0.87 \\
13B & full & 0.97 / 0.96 / 0.86 / 0.68 & 0.42 & 0.49 & 0.97 & 1.00 & 1.21 / 1.54 / 1.02 & 1.17 / 1.49 / 0.99 \\
70B & 16 & 1.08 / 1.09 / 0.98 / 0.89 & 0.41 & 0.44 & 1.09 & 1.12 & 1.10 / 1.23 / 0.99 & 1.07 / 1.20 / 0.96 \\
70B & 32 & 1.10 / 1.10 / 1.01 / 0.89 & 0.42 & 0.46 & 1.11 & 1.14 & 1.05 / 1.16 / 0.98 & 1.01 / 1.12 / 0.95 \\
70B & full & 1.09 / 1.08 / 0.99 / 0.84 & 0.41 & 0.46 & 1.09 & 1.12 & 1.07 / 1.20 / 0.99 & 1.04 / 1.16 / 0.96 \\
\bottomrule
\end{tabular}
\end{table*}

At $p^* = .5$, \sys runs at 0.90--1.02$\times$ of measured self-speculation and 0.87--0.99$\times$ of the rejection-sampling estimate.

\heading{Rejection-sampling self-speculation}
Coupled guessing accepts with probability at most $1 - \mathrm{TV}$, the acceptance of standard speculative sampling, so we estimate rejection-sampling self-speculation from W with $1 - \mathrm{TV}$ measured by a teacher-forced pass over the $\tau = \infty$ outputs (522 documents, about 1.0\,M positions), scaled by the measured over W ratio of the engine's own fixed window.
This is a model-based estimate, and with the coupled agreement in place of $1 - \mathrm{TV}$ it reproduces measured self-speculation within 0.03.

\heading{Batch invariance without exits}
Between $N = 64$ and $N = 1$, 68--70\% of \sys's documents stay identical, against non-EE's 84--88\%, and 23 of its 31 changed documents first differ at a full-depth token, as all of non-EE's do.
Under vLLM's batch-invariant mode, which runs Triton attention in place of FlashInfer, we decode 13B's fidelity cells with \sys, \syscphold, per-row skipping, \syshold without its fallback, and within-iteration batching: all 20,055 decisions compared on identical prefixes agree between $N = 64$ and $N = 1$, with bit-identical ramp margins.

\heading{Where the full-depth fallback ran}
We count the steps on which the fallback (\cref{sec:design-guarantees}) ran the unmodified model, and mark with $\ddagger$ every result where they are at least 1\% of the run's steps.
Compact engines have no fallback, and in-place engines ran it on fewer than 1\% of steps everywhere except here:
\begin{vinlist}
\item synthetic cells run in place: 13B $(.5, 0)$ at every $N$ and 8B $(.5, 0)$ at $N = 16$, where no token would exit, so no token changes, and 8B $(.75, .3)$, $(.25, .5)$, and $(.5, .5)$ at $N = 16$, where 25\%, 44\%, and 44\% of the logged tokens stay involuntarily (\cref{tab:map,tab:map8}), and \syshold at 8B $(.25, .5)$ and $N = 4$ to 16;
\item LayerSkip \syshold: the three 13B quality cells below, sampled PG-19 at $N = 16$ on 13B at $p^* = .5$ and on 8B at every target, and on 8B at $\epsilon^*$ 3\% and $N = 64$ (\cref{tab:layerskip-speed,tab:crossover}), with the 8B $p^* = .5$ and $\epsilon^*$ 1\% twins at $N = 16$, and greedy PG-19 on 13B at $p^* = .5$ after layer 15 (4--7\% of steps).
\end{vinlist}

\heading{\syshold without the full-depth fallback}
The fallback of \syshold turned EE off in three 13B quality cells (MBPP with the exit after layer 15 at $\epsilon^*$ 1\%, and HumanEval at $p^* = .5$ after layers 10 and 15), on 418--438 steps each, so \syshold forced stays there and is not batch-1 EE.
Without the fallback, \syshold exits at batch-1 EE's rate in those cells (.14 / .56 / .49 against .16 / .49 / .45) with its quality, its difference from batch-1 EE contains 0 in 27 of 30 cells and favours it in 3, and \sys minus \syshold without it contains 0 in 29 of 30 cells, the exception against \sys ($-$0.016 [$-$0.032, $-$0.004] on GSM8K after layer 10 at $p^* = .5$).
\syshold without the fallback makes \sys's decisions on 99.76--99.81\% of logged positions.

\section{Draft Heads and the Break-Even Acceptance}
\label{sec:appendix-sd}

\heading{The draft-head model}
We add speculative decoding with a draft head to the calibrated model of \cref{sec:appendix-costmodel}.
A window of $k$ positions runs $k - 1$ draft steps and one verify step, and commits $(1 - \beta^k)/(1 - \beta)$ tokens in expectation at an i.i.d.\ per-position acceptance $\beta$ (vLLM's $k$ drafted tokens are a window of $k + 1$).
A draft step runs one decoder layer of the target's width over $N$ rows with attention over its own KV, the fusion layer that maps $2d$ to $d$, and the target's LM head, inside the verify's engine step, so it pays one layer's fixed cost and no host time of its own.
The verify step runs all layers over $N k$ rows, reading each request's KV prefix once for its $k$ query rows, and the best $k \leq 16$ is taken per $(N, \beta)$.
The model ignores vLLM's own drafting overheads and extra KV, so it is optimistic for the draft head.
In W, a draft step reads 0.018, 0.042, and 0.104 of a full step's weights on 70B, 13B, and 8B.

\heading{Validation}
At each measured cell's tokens per verify, the model predicts EAGLE's speedup with a median measured / predicted ratio of 1.043 on 13B (36 cells), 0.943 on 70B (36), and 1.001 on 8B (136).
Pricing the verify's attention with the reuse factor of \sys's own compacted pass instead under-predicts EAGLE (medians 1.45, 0.99, and 1.25), so vLLM's verify reads each prefix about once.
With shallow layers as the drafter, the same model equals the compact engine's own (\cref{sec:appendix-costmodel}) and predicts the measured self-speculation cells of \cref{sec:appendix-layerskip} within 0.944--1.065 (median 0.984).

\begin{table}[t]
\caption{EAGLE on sampled PG-19 (13B and 70B): EAGLE-1 heads on the Llama-2 chat targets at their best $k$, over non-EE on LayerSkip's checkpoint of the same architecture and lengths, with \sys from the job of that non-EE run (one run, exit after layer 10 on 13B and 20 on 70B, \cref{tab:layerskip-speed} gives three-repeat means), and EAGLE over \sys.}
\label{tab:eagle-pg19}
\centering
\footnotesize
\setlength{\tabcolsep}{3pt}
\begin{tabular}{@{}lr cc cc@{}}
\toprule
 & & & & \multicolumn{2}{c}{\sys at $\epsilon^*$} \\
\cmidrule(lr){5-6}
Model & $N$ & EAGLE & best $k$ & 1\% / 3\% & EAGLE over \sys \\
\midrule
13B & 16 & 1.50 & 3 & 1.14 / 1.39 & 1.32 / 1.08 \\
13B & 64 & 1.52 & 3 & 1.13 / 1.41 & 1.34 / 1.08 \\
13B & 86 & 1.60 & 2 & 1.18 / 1.49 & 1.36 / 1.07 \\
70B & 16 & 1.50 & 3 & 1.19 / 1.34 & 1.26 / 1.12 \\
70B & 32 & 1.49 & 2 & 1.15 / 1.28 & 1.30 / 1.17 \\
70B & 47 & 1.45 & 2 & 1.17 / 1.30 & 1.24 / 1.12 \\
\bottomrule
\end{tabular}
\end{table}

\begin{table}[t]
\caption{Llama-3.1-8B on PG-19 (prompt 201, 2000 tokens), speedup over non-EE at $N = 16$ / 64 / 128 / 256 / 384 / 528 (full memory). Synthetic ramp at $\alpha = .4$ in both modes of \sys, the calibrated model's pick underlined, and per-row skipping and within-iteration batching, each over its own job's non-EE. LayerSkip-8B: exit layer 8, risk targets 3\% and 1\%. $\ddagger$: the full-depth fallback ran (\cref{sec:appendix-layerskip}). --: not run.}
\label{tab:crossover}
\centering
\scriptsize
\setlength{\tabcolsep}{3pt}
\begin{tabular}{@{}l cccccc@{}}
\toprule
System & 16 & 64 & 128 & 256 & 384 & 528 \\
\midrule
\multicolumn{7}{@{}l}{\textit{Sampled}} \\
EAGLE-3, best $k$ & 1.25 & 1.17 & 1.17 & 1.12 & 1.07 & 0.99 \\
(.25, .9) compact & \underline{1.18} & \underline{1.28} & \underline{1.47} & \underline{1.64} & \underline{1.77} & \underline{1.79} \\
(.25, .9) in place & 1.25 & 1.40 & 1.53 & 1.62 & 1.68 & 1.74 \\
(.25, .9) per-row skip. & 0.92 & 1.08 & 1.27 & 1.45 & 1.59 & 1.70 \\
(.25, .9) within-iter. & 0.97 & 1.08 & 1.20 & 1.35 & 1.40 & 1.43 \\
(.25, .5) compact & \underline{1.03} & \underline{1.05} & 1.08 & \underline{1.14} & \underline{1.20} & \underline{1.23} \\
(.25, .5) in place & 0.96 & 1.00 & \underline{1.06} & 1.13 & 1.15 & 1.16 \\
(.25, .5) per-row skip. & 0.91 & 0.98 & 1.08 & 1.15 & 1.20 & 1.24 \\
(.25, .5) within-iter. & 0.95 & 1.01 & 1.07 & 1.14 & 1.15 & 1.16 \\
(.5, .5) compact & \underline{0.94} & \underline{0.95} & \underline{1.02} & \underline{1.07} & \underline{1.10} & \underline{1.12} \\
(.5, .5) in place & 0.96 & 0.98 & 1.02 & 1.06 & 1.07 & 1.08 \\
(.5, .5) per-row skip. & 0.90 & 0.96 & 1.02 & 1.07 & 1.10 & 1.12 \\
(.5, .5) within-iter. & 0.94 & 0.99 & 1.02 & 1.07 & 1.07 & 1.08 \\
LayerSkip-8B 3\% compact & 1.02 & 1.09 & 1.11 & 1.15 & 1.21 & 1.28 \\
LayerSkip-8B 3\% holding only & 1.07$^\ddagger$ & 1.09$^\ddagger$ & 1.05 & 1.10 & 1.14 & 1.16 \\
LayerSkip-8B 1\% compact & 0.97 & 0.97 & 1.01 & 1.03 & 1.09 & 1.07 \\
LayerSkip-8B 1\% holding only & 0.97$^\ddagger$ & 0.98 & 1.00 & 1.04 & 1.07 & 1.08 \\
LayerSkip-8B 3\% per-row skip. & 0.92 & 1.00 & -- & 1.12 & -- & 1.27 \\
LayerSkip-8B 3\% within-iter. & 0.94 & 1.01 & 1.06 & 1.11 & 1.14 & 1.17 \\
LayerSkip-8B 1\% per-row skip. & 0.92 & 0.96 & -- & 1.02 & -- & 1.09 \\
LayerSkip-8B 1\% within-iter. & 0.94 & 0.99 & 1.00 & 1.04 & 1.07 & 1.07 \\
\midrule
\multicolumn{7}{@{}l}{\textit{Greedy}} \\
EAGLE-3, best $k$ & 2.26 & 1.90 & 1.61 & 1.43 & 1.36 & 1.32 \\
(.25, .9) compact & \underline{1.30} & \underline{1.38} & \underline{1.52} & \underline{1.69} & \underline{1.81} & \underline{1.93} \\
(.25, .9) in place & 1.32 & 1.40 & 1.50 & 1.68 & 1.74 & 1.80 \\
(.25, .9) per-row skip. & 0.94 & 1.09 & 1.30 & 1.57 & 1.76 & 1.89 \\
(.25, .9) within-iter. & 0.96 & 1.10 & 1.23 & 1.43 & 1.49 & 1.53 \\
(.25, .5) compact & \underline{1.08} & \underline{1.09} & 1.12 & \underline{1.19} & \underline{1.28} & \underline{1.32} \\
(.25, .5) in place & 0.97 & 1.00 & \underline{1.07} & 1.17 & 1.20 & 1.22 \\
(.25, .5) per-row skip. & 0.92 & 0.98 & 1.09 & 1.20 & 1.28 & 1.34 \\
(.25, .5) within-iter. & 0.94 & 1.01 & 1.08 & 1.18 & 1.21 & 1.22 \\
(.5, .5) compact & \underline{1.01} & \underline{1.01} & \underline{1.03} & \underline{1.10} & \underline{1.16} & \underline{1.19} \\
(.5, .5) in place & 0.98 & 0.99 & 1.03 & 1.10 & 1.12 & 1.13 \\
(.5, .5) per-row skip. & 0.92 & 0.96 & 1.03 & 1.10 & 1.16 & 1.19 \\
(.5, .5) within-iter. & 0.94 & 0.99 & 1.04 & 1.10 & 1.12 & 1.13 \\
\bottomrule
\end{tabular}
\end{table}

\heading{The 8B grid}
Each system is measured against non-EE in its own job, and the non-EE runs agree within 1.4\% on the cells they share (\cref{tab:crossover}).
EAGLE-3's best $k$ is 1 on every sampled cell and falls from 6 to 1 or 2 with greedy decoding, at a per-position acceptance of .35--.38 sampled and .72--.75 greedy.
On 8B the calibrated choice between compact and in-place mode can be wrong below full memory, because no 8B shallow step entered the fit: in place runs up to 9\% faster than the compact pick with sampled text at $N \leq 64$.
The deployed numbers in this paper keep the calibrated choice.

\heading{Scope}
Draft heads here draft chains, and trees~\cite{eagle2} raise tokens per verify where rows are free, which favours them at small $N$.
Our contexts are about 2K tokens, and with long contexts large batches become bound by KV reads, where speculative decoding pays again~\cite{magicdec2025}.

\end{document}